\documentclass[3p,times]{elsarticle}

\usepackage{ecrc}

\volume{00}

\firstpage{1}

\journalname{Physica A}

\runauth{P.~Jizba and J.~Korbel}

\jnltitlelogo{Physica A}

\CopyrightLine{2014}{Published by Elsevier Ltd.}

\usepackage{amssymb}
\usepackage{amsthm}

\usepackage[usenames,dvipsnames]{color}
\usepackage{amsmath}
\usepackage{dsfont}
\usepackage{graphicx}
\usepackage{algorithm}
\usepackage{float}
\usepackage{algpseudocode}
\usepackage{multirow}
\newfloat{algorithm}{t}{lop}
\newcommand{\ud}{\mathrm{d}}
\newcommand{\card}{\mathrm{card}}
\newcommand{\E}{\mathbb{E}}

\definecolor{darkred}{rgb}{.8,0,0}

\definecolor{darkblue}{rgb}{0,0,.7}

\begin{document}

\begin{frontmatter}



\dochead{}

\title{Multifractal Diffusion Entropy Analysis: Optimal Bin Width of Probability
Histograms}


\author[FNSPE,FU]{Petr Jizba}
\ead{p.jizba@fjfi.cvut.cz}

\author[FNSPE,MPI]{Jan Korbel}
\ead{korbeja2@fjfi.cvut.cz}

\address[FNSPE]{Faculty of Nuclear Sciences and Physical Engineering, Czech Technical University in Prague, B\v{r}ehov\'{a} 7,
11519, Prague, Czech Republic}
\address[FU]{Institute of Theoretical Physics, Freie Universit\"{a}t in Berlin, Arnimallee 14, 14195 Berlin, Germany}
\address[MPI]{Max Planck Institute for the History of Science, Boltzmannstrasse 22, 14195 Berlin, Germany}

\begin{abstract}
In the framework of Multifractal Diffusion Entropy Analysis we
propose a method for choosing an optimal bin-width in histograms
generated from underlying probability distributions of interest. The
method presented uses techniques of R\'{e}nyi's entropy
and the mean squared error analysis to discuss the conditions under which the
error in the multifractal spectrum estimation is minimal. We illustrate
the utility of our approach by focusing on a scaling behavior of
financial time series. In particular, we analyze the S\&P500 stock
index as sampled at a daily rate in the time period 1950-2013. In
order to demonstrate a strength of the method proposed we compare
the multifractal $\delta$-spectrum for various bin-widths and show
the robustness of the method, especially for large values of $q$. For such values, other methods
in use, e.g., those based on moment estimation, tend to fail for heavy-tailed data or data with long
correlations. Connection between the $\delta$-spectrum and
R\'{e}nyi's $q$ parameter is also discussed and elucidated on a simple example of
multiscale time series.

%
%
\end{abstract}

\begin{keyword}
Multifractals \sep R\'{e}nyi entropy \sep Stable distributions \sep Time series
\PACS 89.65.Gh \sep 05.45.Tp

\end{keyword}

\end{frontmatter}



\section{Introduction}\label{Sec1}

The evolution of many complex systems in natural, economical, medical and biological sciences is usually
presented in the form of time data-sequences. A global massification of computers together with their improved ability
to collect and process large data-sets have brought about the need for novel analyzing methods.
A considerable amount of literature has been recently devoted to developing and using new data-analyzing paradigms.
These studies include such concepts as fractals and multifractals~\cite{Kim:04},  fractional dynamics~\cite{Machado,West}, complexity~\cite{Park,Lee:physics0607282}, entropy densities~\cite{Lee:physics0607282} or transfer entropies~\cite{Schreiber,Marschinski,transferent}.
Particularly in the connection with financial time series there has been rapid development of
techniques for measuring and managing the fractal and multifractal scaling behavior from
empirical high-frequency data sequences.
A non-trivial scaling behavior in a time data-set represents a typical signature of a multi-time scale cooperative behavior in much
the same way as a non-trivial scaling behavior in second-order phase transitions reflects the underlying long-range (or multi-scale) cooperative interactions. The usefulness of the scaling approach is manifest, for instance, in quantifying critical or close-to-critical scaling which typically signalizes onset of financial crises, including stock market crashes, currency crises or sovereign defaults~\cite{kleinert:a}. A multifractal scaling, in particular, is instrumental in
identifying the relevant scales that are involved in both temporal and inter-asset correlations~\cite{transferent}.
In passing, one can mention that aside from {\em financial} data sequences, similar (multi)fractal scaling patterns are also observed (and analyzed)
in time data-sets of
heart rate dynamics~\cite{Peng2,Voit}, DNA sequences~\cite{Peng,mantegna}, long-time
weather records~\cite{Talker:00}  or in electrical power loads~\cite{Bozic:13}.


In order to identify fractal and multifractal scaling in time series generated by a complex system (of both deterministic and stochastic nature),
several tools have been developed over the course of time. To the most popular ones belong
the Detrended Fluctuation Analysis~\cite{Peng,mfdfa}, Wavelets~\cite{multiwavelets}, or Generalized Hurst Exponents~\cite{generhurst}. The purpose of the present paper is to discuss and advance
yet another pertinent method, namely the Multifractal Diffusion Entropy Analysis
(MF-DEA). In doing so we will stress the key r\^{o}le that R\'{e}nyi's entropy (RE) plays in this context. To this end,
we will employ two approaches for the estimation of the scaling exponents that can be directly phrased in terms of RE, namely,
the monofractal approach of Scafetta {\em et al.}~\cite{dea} and the multifractal approach
of Huang {\em et al.}~\cite{mdea}, with further comments of Morozov~\cite{commentmfdea}. The most important upshot that will emerge from this study is
the proposal for the optimal bin-width in empirical histograms. The latter ensures that the error in the RE
(and hence also scaling exponents) evaluation, when the underlying probability density function (PDF) is replaced by its empirical histograms,
is minimal in the sense of R\'{e}nyi's information divergence and the associated $L_2$-distance.
We further show that the ensuing optimal bin-width permits the
characterization of the hierarchy of multifractal scaling exponents $\delta(q)$ and $D(q)$ in a fully quantitative fashion.

This paper is structured as follows: In Section~\ref{Sec2} we
briefly review foundations of the multifractal analysis that will be
needed in following sections. In particular, we introduce such
concepts as Lipschitz--H\"{o}lder's singularity exponent,
multifractal spectral function and their Legendre conjugates. In
Section~\ref{Sec3} we state some fundamentals of the Fluctuation
collection algorithm and propose a simple instructive example of a
heterogeneous multiscale time series. Within this framework we
discuss the MF-DEA and highlight the r\^{o}le of R\'{e}nyi's entropy
as a multiscale quantifier. After this preparatory material we turn
in Section~\ref{sec.4} to the question of the optimal bin-width
choice that should be employed in empirical histograms. In
particular, we analyze the bin-with that is needed to minimize error
in the multifractal spectrum evaluation. In Section~\ref{Sec.5}, we demonstrate the
usefulness and formal consistency of the proposed error estimate by
analyzing time series from S\&P500 market index  sampled at a daily
(end of trading day) rate basis in the period from January 1950 to
March 2013 (roughly 16000 data points).
We apply the symbolic computations with the open source software $R$ to illustrate the strength of the proposed optimal
bin-width choice. In particular, we graphically compare the multifractal $\delta$-spectrum for various bin-widths.
Our numerical results imply that the proposed bin-widths are indeed optimal in comparison with other alternatives used.
Implications for the $\delta(q)$-spectrum as a function of R\'{e}nyi's $q$ parameter are also discussed and graphically represented.
Conclusions and further discussions are relegated to the concluding section. For the reader's convenience, we present in~\ref{sect: appendix}
the source code in the language $R$ that can be directly employed for efficient estimation of the  $\delta(q)$-spectrum
(and ensuing generalized dimension $D(q)$) of long-term  data sequences.

\section{Multifractal analysis\label{Sec2}}

Let us have a discrete time series $\{x_j\}_{j=1}^N \subset \mathds{R}^D$, where $x_j$ are obtained from measurements at times $t_j$ with an equidistant
time lag $\mathfrak{s}$. We divide the whole domain of definition of $x_j$'s into distinct regions $K_i$ and define the probability of each region as
\begin{equation}
p_i \ \equiv \ \lim_{N \rightarrow \infty} \frac{N_i}{N} \ = \
\lim_{N \rightarrow \infty} \frac{\card\{j \in \{1, \dots, N\} |\ \!
x_j \in K_i \}}{N}\, ,\label{2.1a}
\end{equation}
where ``card'' denotes the {\em cardinality}, i.e.,  the number of elements contained in a set. For every region, we consider that the probability
scales as $p_i(\mathfrak{s}) \propto \mathfrak{s}^{\alpha_i}$, where $\alpha_i$ are scaling exponents also known as the Lipschitz--H\"{o}lder (or singularity) exponents. The key assumption in the multifractal analysis is that in the small-$\mathfrak{s}$  limit we can assume that the probability distribution depends {\em smoothly} on $\alpha$ and thus the probability that some arbitrary region
has the scaling exponent in the interval $(\alpha, \alpha + \ud \alpha)$ can be considered in the form
\begin{equation}
 \ud \rho(\mathfrak{s},\alpha) \ = \ \lim_{N \rightarrow \infty} \frac{  \card \{p_i \propto  \mathfrak{s}^{\alpha'}|\ \! \alpha'
  \in (\alpha, \alpha + \ud \alpha) \} }{N} \ = \ c(\alpha) \mathfrak{s}^{-f(\alpha)} \ud \alpha \, .\label{2.2a}
\end{equation}
The corresponding scaling exponent $f(\alpha)$ is known as the {\em multifractal spectrum} and by its very definition it represents the (box-counting) fractal dimension of the subset that carries PDF's with the scaling exponent $\alpha$.

A convenient way how to keep track with various $p_i$'s is to examine the scaling of the correspondent moments. To this end one can define a ``partition function''
\begin{equation}\label{eq: partition}
Z(q,\mathfrak{s})\ = \ \sum_i p_i^q \ \propto \ \mathfrak{s}^{\tau(q)}\, .
\end{equation}
Here we have introduced the {\em scaling function} $\tau(q)$ which  is (modulo a multiplicator) an analogue of thermodynamical free energy~\cite{Renyi,stanley}. In its essence is the partition function (\ref{eq: partition}) nothing but the $(q-1)$-th order moment of the probability distribution\footnote{$\mathcal{P}^{q-1}$ is a random variable with probabilities $p_i$ and values equal to $p_i^{q-1}$.}, i.e.,
$Z(q,\mathfrak{s}) = \E[\mathcal{P}^{q-1}(\mathfrak{s})]$. It is sometimes convenient to introduce the generalized mean of the random variable
$\mathcal{X} = \{x_i\}$  as
\begin{equation}
 \E[\mathcal{X} ]_{f} = f^{-1} \left(\sum_i f(x_i) p_i \right) \, .
\end{equation}
The function $f$ is also known as the Kolmogorov--Nagumo function~\cite{Renyi}. For the choice $f(x) = x^{q-1}$ one  obtains the so-called $q$-mean which is typically denoted as $\E[ \cdots ]_q$. It is then customary to introduce the scaling exponent $D(q)$ of the $q$-mean as
\begin{equation}
\E[ \mathcal{P}(\mathfrak{s}) ]_q \ = \ \sqrt[q-1]{\E[ \mathcal{P}^{q-1}(\mathfrak{s})]} \ \propto \ \mathfrak{s}^{D(q)} \, .
\end{equation}
Such $D(q)$ is called a {\em generalized dimension} and from~\eqref{eq: partition} it is connected with $\tau(q)$ via the relation: $D(q) = {\tau(q)}/{(q-1)}$. In some specific situations, the generalized dimension can be identified with other
well known fractal dimensions, e.g., for $q=0$ we have the usual {\em box-counting fractal dimension}, for $q \rightarrow 1$ we have the {\em informational dimension} and for $q=2$ it corresponds to the {\em correlation dimension}. The generalized dimension  itself is obtainable from the relation
\begin{equation}
D(q) \ = \ \lim_{\mathfrak{s} \rightarrow 0} \frac{1}{q-1} \frac{\ln Z(q,\mathfrak{s})}{\ln \mathfrak{s}} ,
\label{2.6.aa}
\end{equation}
which motivates the introduction of  R\'{e}nyi's entropy\footnote{Here and throughout we use the natural logarithm, thought from the information point of view it is more natural to phrase RE via the logarithm to the base two. RE thus defined is then measured in natural units --- nats, rather than bits.}
\begin{eqnarray}
H_q(\mathfrak{s}) \ = \ \frac{1}{1-q}\ln Z(q,\mathfrak{s})\, .
\label{RE1}
\end{eqnarray}
The generalized dimension defined via (\ref{2.6.aa}) is also known as a {\em R\'{e}nyi dimension} of order $q$.
Eq.~(\ref{RE1}), in turn, implies that for small $\mathfrak{s}$ one has $H_q(\mathfrak{s}) \sim - D(q) \ln \mathfrak{s} + C_q$ where $C_q$ is the term independent of $\mathfrak{s}$ (cf. Ref.~\cite{Renyi}).
The multifractal spectrum $f(\alpha)$ and scaling function $\tau(q)$ are not independent. The relation between them can be found from the expression for the partition function which can be, with the help of (\ref{2.1a})  and (\ref{2.2a}), equivalently written as
\begin{equation}
Z(q,\mathfrak{s}) \ = \ \int \ud \alpha \ \! c(\alpha) \mathfrak{s}^{-f(\alpha)} \mathfrak{s}^{q \ \!\alpha} \, .
\end{equation}
By taking the the steepest descent approximation one finds that
\begin{equation}\label{eq: leg}
\tau(q) \ = \ q \alpha(q) - f(\alpha(q)) \, .
\end{equation}
Here $\alpha(q)$ is such a value of the singularity exponent, that maximizes $q \alpha - f(\alpha)$ for given $q$. Together with the assumption of differentiability, one finds that  $\alpha(q) = {\ud \tau(q)}/{\ud q}$, and hence Eq.~\eqref{eq: leg} is nothing but the Legendre transform between two conjugate pairs $(\alpha,f(\alpha))$ and $(q,\tau(q))$. The Legendre transform $f(\alpha)$ of the function $\tau(q)$ contains the same information as $\tau(q)$, and is the characteristic customarily studied when dealing with multifractals~\cite{multiwavelets,mdea,Renyi,stanley}.

In passing we can observe that the passage from multifractals to
monofractals is done by assuming that the $\alpha$-interval gets
infinitesimally narrow an the underlying PDF is smooth. In such a
case both $\alpha$ and $f(\alpha)$ collapse to $\alpha = f(\alpha)
\equiv D$ and $q = {\ud f(\alpha)}/{\ud \alpha} = 1$ (cf.
Ref.~\cite{Renyi}).

\section{Diffusion entropy analysis\label{Sec3}}

\subsection{Fluctuation collection algorithm\label{Sec3a}}

In order to be able to utilize RE for extracting scaling exponents, we need to
correctly estimate probability distribution from given empirical data sequence.
The method that is particularly useful for our purpose is the so-called {\em
Fluctuation collection algorithm} (FCA)~\cite{dea}.
To understand what is involved, let us assume that $\{x_j\}_{j=1}^N$
is a {\em stationary} time series. This can be converted to a random
walk (i.e., a diffusion trajectory) by considering the time series
as being a set of one-dimensional diffusion-generating {\em fluctuations}. With
this a diffusing ``particle's'' position at time $t$ will be $\xi(t)
= \sum_{i=1}^M x_i$.
%
Here $t = M\mathfrak{s}$ with $M \leq N$ and $\mathfrak{s}$ being the elementary time lag.
The reason behind usage of random walk-like process instead of the
original (noise-like) data time series is the Central Limit Theorem
(CLT) and its L\'{e}vy--Gnedenko heavy-tailed
generalization~\cite{mantegna,Kolmogorov}. In fact, the ensuing
distributions in the CLT basin of attraction allow to identify (and
quantify) the truly universal scaling behavior imprinted in the auxiliary
diffusion process. This is because the non-trivial scaling can be
phrased in terms of parameters of the involved
L\'{e}vy stable non-Gaussian distributions.

As a rule, the inter-time scale correlations typically cause the
breakdown of the CLT globally but in many cases  the
validity of the CLT is retained at least locally. In such a case the L\'{e}vy
scaling parameters depend on the local time scale, which in
turn leads to a typical multifractal picture seen in time
series born out of complex dynamical systems. An example of this
kind of behavior will be presented in the next subsection.

%
In order to establish the possible existence of a scaling
we need to know the probability for the auxiliary diffusion trajectory to be found at the position $x$
at time $s = n \mathfrak{s}$ ($n$ an integer satisfying the condition $1 \leq n \leq N$).
The FCA addresses precisely this issue by turning a single
time sequence, or better the derived diffusion trajectory  into
a Gibbs ensemble of diffusion trajectories. To this end one
defines $N-n +1 $  sub-sequences $\{ x_j^{(\kappa)} \}_{j=0}^{n}$ where
\begin{eqnarray}
x_j^{(\kappa)} \ &=& \ x_{j+\kappa}, \;\;\;\;\;\; \kappa \ = \ 0, \ldots, M-n\, , \;\;\;\;\; j = 1, \dots, n\, , \nonumber \\
x^{(\kappa)}_0 \ &\equiv& \ 0, \;\;\;\;\;\;\;\;\;\; \kappa \ = \ 0, \ldots, M-n\, .
\end{eqnarray}
In particular, each aforementioned sub-sequence has the time duration $s$. One might envisage
that the sub-sequences with different $\kappa$ are formed by shifting
a window of a fixed length $s$ by a constant amount $\mathfrak{s}$ within the original time sequence.
There will be clearly only  $N-n +1$ such shifts possible. For this reason is such a selection
of sub-sequences also known as the  {\em mobile window} method~\cite{Grigolini}.

\begin{figure}[t]
\begin{center}
\includegraphics[width=10cm]{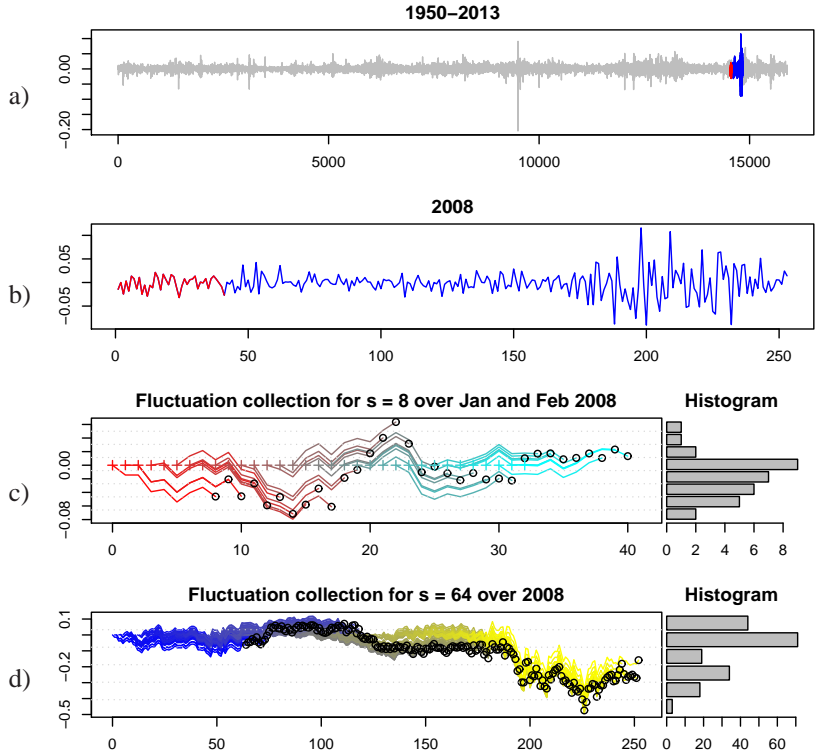}
\caption{Illustration of fluctuation collection algorithm. From
above: a) Time series of financial index S\&P500 from January 1950
to March 2013, containing approx. 16000 entries. b) A subset of
S\&P500 for the year 2008 alone. c) Fluctuation collection algorithm
for the first two months of 2008 and $s=8$ ($\mathfrak{s}$ is $1$
day). The series is partially integrated, i.e., fluctuation sums
$\sigma_8(t)$ (defined in Eq.~\eqref{eq: sigma}) are collected into
the histogram on the right-hand-side. d) Fluctuation collection
algorithm for the whole year 2008 for $s=64$ ($\mathfrak{s}$ is $1$
day). This histogram was
estimated independently of the first histogram (for every histogram was separately applied the Scott optimal bin-width rule, broadly discussed in following sections), and therefore
estimated bin-widths for both histogram differ. This is a problem
for RE estimation, so it is necessary to find some common bin-width
for all scales.}
\label{fig: fluct}
\begin{picture}(20,7)
\put(-100,382){\small{Fluctuation collection algorithm for time series S\&P
500}}
\put(-150,340){a)}
\put(-150,265){b)}
\put(-150,190){c)}
\put(-150,120){d)}
\end{picture}
\end{center}
\end{figure}

From any sub-sequence $\{x_j^{(\kappa)} \}_{j=0}^{n}$ one may form an auxiliary diffusion trajectory
(basically a sample path) as
\begin{equation}\label{eq: sigma}
\xi^{(\kappa)}(s) \ = \ \sum_{j=0}^n x^{(\kappa)}_{j}\ = \ \sum_{j=0}^n x_{j+\kappa}\, .
\end{equation}
%
This mapping of a time series to a diffusion process generates  an overlapping diffusion process.
The ensemble of such $N-s +1$ diffusion trajectories allows now to find the required probability;
all obtained values of $\xi^{(\kappa)}(s)$ for different $\kappa$'s are divided into regions $K_i$ of length $h(s)$ and the
probability is estimated from the (normalized) equidistant histogram, i.e.
\begin{equation}
\hat{p}_i(s) \ \equiv \ \frac{\card\{t\ \!| \ \! \xi_s(t) \in  K_i
\}}{N-n+1} \, .
\end{equation}
The aforementioned FCA is illustrated in Fig.~\ref{fig: fluct}  with the
time series of the S\&P500 index.
In case of multidimensional data series, i.e., when $\{x_j\}_{j=1}^N
\subset \mathds{R}^D$, we estimate $D$-dimensional histogram with
hyper-cubical bins of the elementary volume $h^D$. Unfortunately as it stands, the FCA does not prescribe the size of $h(s)$ which is, however, crucial for the correct RE evaluation.
In the following subsection we will see that often $s = s(q)$ and hence the optimal
bin width $h$ will be generally $q$-dependent. In Section~\ref{sec.4} we will discuss the actual optimal value of such
$h(q)$.
%

It is clear that the FCA is not sensibly applicable in cases of
non-ergodic and non-Markovian systems, phrased, e.g., via
accelerating, path-dependent or aging random walks. This is because
the key assumption behind the whole fluctuation collection
construction is a reliable use of the Gibbs ensemble method which,
in turn is justified only when the time series in question
represents a time evolution of an underlying system, whose complex
statistical rules do not change in time. It will be this limited
framework of the FCA that will be, for simplicity's sake, utilized
in sections to follow. Despite its limited applicability, one is
still able to draw from this scheme valuable conclusions about a scaling behavior of
complex multi-scale data sets. Interested reader may find some
generalizations of the FCA to non-stationary time series, e.g.,
in Refs.~\cite{Grigolini, GrigoliniII}.

\subsection{Monofractal and Multifractal diffusion entropy analysis\label{Sec3b}}

Statistical fractals and multifractals cannot be described comprehensively by
descriptive statistical measures, such as mean or variance, because these do
depend on the scale of observation in a power law fashion.
In the literature one can find a number of approaches allowing to extract above
multifractal scaling exponents from empirical data sequences. To these belong,
for instance, Detrended Fluctuation Analysis constructed on random
walks~\cite{Peng,mfdfa}, signal theory based Wavelet
Analysis~\cite{multiwavelets}, Generalized Hurst Exponents~\cite{generhurst},
etc. In the following we will employ yet two another approaches for the
estimation of scaling exponents that are directly based on the use of RE,
namely, the Monofractal Diffusion Entropy Analysis of Scafetta {\em et
al.}~\cite{dea} and the MF-DEA of Huang {\em et al.}~\cite{mdea} and
Morozov~\cite{commentmfdea}. The reason for choosing these approaches is that
unlike other methods in use, the diffusion entropy analysis can determine the
correct scaling exponents even when the statistical properties are anomalous or
strange dynamics (e.g., long-range patterns in evolution) are involved.

Many techniques in modern finance rely substantially on the assumption that the
random variables under investigation follow a Gaussian distribution. However,
time series observed in finance --- but also in other applications --- often deviate
from the Gaussian model, in that their marginal distributions are heavy-tailed
and often even asymmetric. In such situations, the appropriateness of the commonly
adopted normal assumption is highly questionable.
It is often argued that financial asset returns are the cumulative outcome of
a large number of pieces of information and individual decisions arriving almost
continuously in time. Hence, in the presence of heavy-tails it is natural
to assume that they are approximately governed by some (L\'{e}vy) stable non-Gaussian distribution.
Other leptokurtic distributions, including Student's $t$, Weibull, and
hyperbolic, lack the attractive CLT property. RE can deal very efficiently with heavy-tailed
distributions that occur in real-world complex dynamics epitomized, e.g., in biological
or financial time series~\cite{transferent}. Moreover, as we have already seen,
RE is instrumental in uncovering a self-similarity present in
the underlying distribution~\cite{beck}. To better appreciate the r\^{o}le of RE
in a complex data analysis and to develop our study further,
it is helpful to examine some simple model situation. To this end, let us begin
to assume
that the PDF $p(x,t)$ has the self-similarity encoded via a simple scaling
relation~\cite{Grigolini}
\begin{equation}
p(x,t)  \ = \ \frac{1}{t^\delta} F\left(\frac{x}{t^\delta}\right)  \, ,
\label{3.10.aa}
\end{equation}
with $t$ being a dimensionless integer (say $N \geq 1$) measuring elapsed time by counting
timer ticks that are $\mathfrak{s}$ units apart. Alternatively we could
use in (\ref{3.10.aa}) instead of the (dimensionless) $t$ the fraction $t/\mathfrak{s}$ with both $t$ and
$\mathfrak{s}$ dimension full. In the latter case $t$ and  $\mathfrak{s}$ would be measured in seconds.

Such a scaling is known to hold, for instance, in Gaussian distribution,
exponential distribution or more generally for L\'{e}vy stable distributions.
For stable distributions, the relation (\ref{3.10.aa}) basically means that the
probability $p(x,t)$, when understood as a sum of $N$ random
variables
is, up to a certain  scale factor, identical to that of individual
variables, i.e., $F(x)$.
This is a typical signature of fractal behavior
--- the whole looks like its parts. One way how to operationally extract the
scaling coefficient $\delta$ is to employ the differential (or continuous)
Shannon entropy
\begin{equation}
H_1(t) \ = \ - \int_{\mathds{R}} \ud x \, p(x,t) \ln [p(x,t)]\, ,
\end{equation}
because in such a case
\begin{eqnarray}
H_1(t) \ = \ A + \delta \ln t\, ,
\end{eqnarray}
with $A$ being a $t$-independent constant. So $\delta$ can be decoded from
log-lin plot in the ($t, H_1$) plane. Note that for the  Brownian motion this
gives $\delta = 1/2$ which implies the well-known Brownian scaling, and for
(L\'{e}vy) stable distribution is the $\delta$-exponent related with the
L\'{e}vy $\mu$ parameter via relation $\delta = 1/\mu$.

Although the above single-scale models capture much of the essential dynamics
involved, e.g., in financial markets, they neglect the phenomenon of temporal
correlation that in turn leads to both long- and short-range heterogeneities
encountered in empirical time sequences.  Such correlated time series typically
do not obey the central limit theorem and  are described either with anomalous
single-scaling exponent or via multiple-scaling. As a rule, in realistic time
series the scaling exponents differ at different time scales, i.e., for each
time window of size\footnote{Here $s$ is again a dimensionless (time) tick count variable, i.e., {\em integer}.}
$s$ (within some fixed time horizon $t$) exponents
$\delta(s)$ have generally different values for different  $s$. In order to avoid technical
difficulties, let us consider that $t = m \cdot s$ such that $m \in \mathds{N}$ which means,
that $N$ is divisible by $m$. One can further
assume that for each fixed time scale $s$ the underlying processes are
statistically independent and so the total PDF with the time horizon $t$  can be
written as a convolution of respective PDF's, i.e.
%
%
\begin{equation}
p(x,s,t) \ = \ \int  \, \ud^{m} {\mathbf z}  \ \,
\delta\left(x - \sum_{i=1}^{m} z_i\right)
\prod_{j=1}^{m} \ \! \frac{1}{s^{\delta(s)}} \
\!F\left(\frac{z_i}{s^{\delta(s)}} \right) \, ,
\label{3.16.aac}
\end{equation}
where
Note in particular that (\ref{3.16.aac}) can be equivalently rewritten as
\begin{equation}
p(x,s,t) \ = \  \frac{1}{s^{\delta(s)}} \int  \, \ud^{m} {\mathbf z}  \ \,
\delta\left(\frac{x}{s^{\delta(s)}} - \sum_{i=1}^{m} z_i\right)
\prod_{j=1}^{m}  \
\!F\left({z_i} \right)  \ \equiv \ \frac{1}{s^{\delta(s)}} G\left(\frac{x}{s^{\delta(s)}}\right) \, ,
\end{equation}
where
\begin{eqnarray}
G(x) \ = \ (\underbrace{F\ast F \ast \cdots \ast F}_{m~\rm{times}})(x)\, .
\end{eqnarray}

In order to obtain a non-trivial scale-dependent scaling we will assume
that the PDF $F(x)$ is the L\'{e}vy stable distribution. For simplicity's sake we will consider only
the symmetric $\mu$-stable L\'{e}vy distribution, i.e., distribution of the form
\begin{eqnarray}
L_{s,\ \!\mu}(x) \ = \ \frac{1}{2\pi}\int_{\mathds{R}} \ud k \ \! \exp\left(-s|k|^{\mu(s)}\right) \ \! e^{-i k x} \ = \
\frac{1}{\pi}\int_{0}^{\infty} \ud k \ \! \cos(kx) \ \! \exp\left(-sk^{\mu(s)}\right)
\ = \ \frac{1}{s^{1/\mu(s)}}L_{1,\ \!\mu}\left(\frac{x}{s^{1/\mu(s)}}\right)\, ,
\label{3.19.abc}
\end{eqnarray}
where $s> 0$ and $\mu(s) \in (0,2]$ are the L\'{e}vy {\em scale parameter} and L\'{e}vy {\em stability index}, respectively.
This allows to identify $\delta(s)$ with $1/\mu(s)$ and $F(x)$ with $L_{1,\ \!\mu}(x)$.

Analogously as in the single-scale case, we can pin-point a concrete scale-dependent scaling by
looking at the dependence of the differential entropy on the time scale $s$.
As mentioned in Section~\ref{Sec2}, the monofractals correspond to situations with $q=1$,
which is also the basic reason
why the RE of order $1$ (i.e., Shannon's entropy) plays such an important r\^{o}le in the monofractal DEA. On the other hand,
in the multifractal case (such as the case of scale dependent distribution (\ref{3.16.aac}))  we should use
instead of the Shannon differential entropy the whole class of
differential RE, defined as
\begin{equation}
H_q(s,t) \ = \ \frac{1}{1-q} \ln \!\int_{\mathds{R}} \ud x \, p^q(x,s,t)\, .
\label{3.15.b}
\end{equation}
to obtain the local scaling $\delta(s)$. Indeed, by plugging (\ref{3.16.aac}) with (\ref{3.19.abc}) to (\ref{3.15.b})
we receive
\begin{eqnarray}
H_q(s,t) \ &=& \  \frac{1}{\mu(s)} \ln s \ + \ \frac{1}{1-q} \ln \! \int_{\mathds{R}} \ud x \ G^q(x) \ = \
\frac{1}{\mu(s)} \ln s \ + \ \frac{1}{1-q} \ln \! \int_{\mathds{R}} \ud x \ L^q_{m, \ \!\mu}(x)\nonumber\\[2mm]
&=& \ \frac{1}{\mu(s)} \ln t \ + \ \frac{1}{1-q} \ln \! \int_{\mathds{R}} \ud x \ L^q_{1, \ \!\mu}(x)\, .
\label{3.21.abc}
\end{eqnarray}
As a consequence of the asymptotic behavior of $L^q_{1, \ \!\mu}(x)$ for $\mu < 2$ (which at large  $|x|$ behaves as $1/|x|^{\mu+1}$), the formula (\ref{3.21.abc}) has a good mathematical sense only for $q > 1/(1+\mu)$. The limit case $\mu =2$ corresponds to the Gaussian distribution which is well behaved for any $q>0$. We may note further that the scale $s$ is not explicitly present in (\ref{3.21.abc}), instead it is implicitly reflected only via the stability index $\mu(s)$.
This is understandable given that the scale $s$ cannot be directly observed, while $H_q$ is, in principle, an observable quantity~\cite{JA2,Bashkirov:06}. When the time lag $\mathfrak{s}$ is small one might expect that $H_q$ can be analytically extended from
$s\in \mathbb{N}^+$ to $s\in \mathbb{R}^+$.
It is then easy to check that
\begin{eqnarray}
\label{eq: Hq}
0 \ = \ \frac{\partial H_q(s,t)}{\partial s}  \ = \ - \mu'(s)\left[\frac{\ln t}{\mu(s)^2 } \ - \frac{q}{1-q} \ \frac{\int_{\mathds{R}} \ud x \  (\partial L_{1, \ \!\mu}(x)/ \partial \mu )  \, L^{q-1}_{1, \ \!\mu}(x) }{\int_{\mathds{R}} \ud x \ L^q_{1, \ \!\mu}(x)  }   \right]\, ,
\end{eqnarray}
and, in particular
\begin{eqnarray}
0 \ = \ \left.\frac{\partial H_q(s,t)}{\partial s} \right|_{q \rightarrow 1} \ &=& \ - \mu'(s)\left[\frac{\ln t}{\mu(s)^2 } \ +  \ \int_{\mathds{R}} \ud x \  (\partial L_{1, \ \!\mu}(x)/ \partial \mu )  \, \left(\ln L_{1, \ \!\mu}(x) + 1 \right)\right]\nonumber \\[3mm]
&=& \ - \mu'(s)\left[\frac{\ln t}{\mu(s)^2 } \ +  \ \frac{\partial}{\partial \mu }\int_{\mathds{R}} \ud x \   L_{1, \ \!\mu}(x)\ \! \ln L_{1, \ \!\mu}(x)\right]
\, .
\label{3.24aaa}
\end{eqnarray}
%
%

\begin{figure}[t]
\begin{center}
\includegraphics[width=7cm]{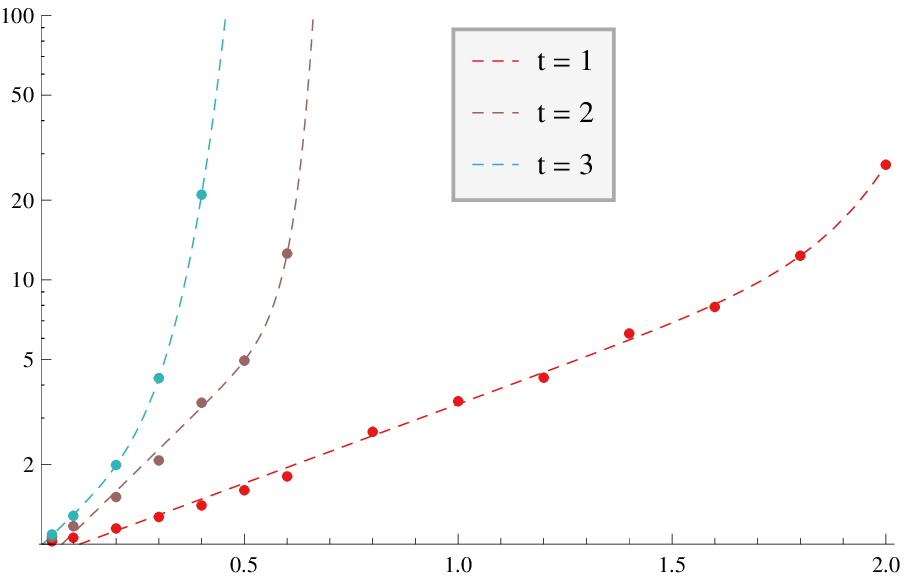} \quad \quad
\includegraphics[width=7cm]{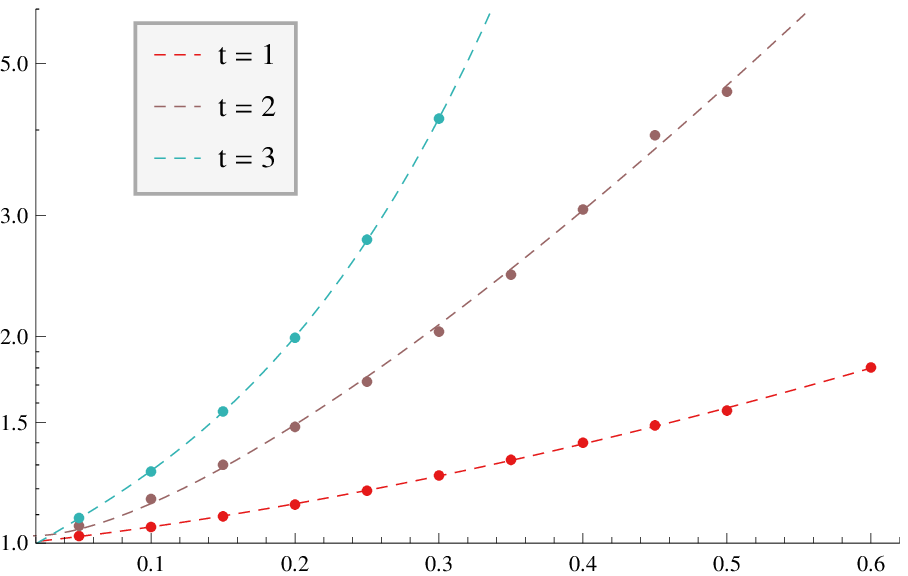}\\[8mm]
\includegraphics[width=7cm]{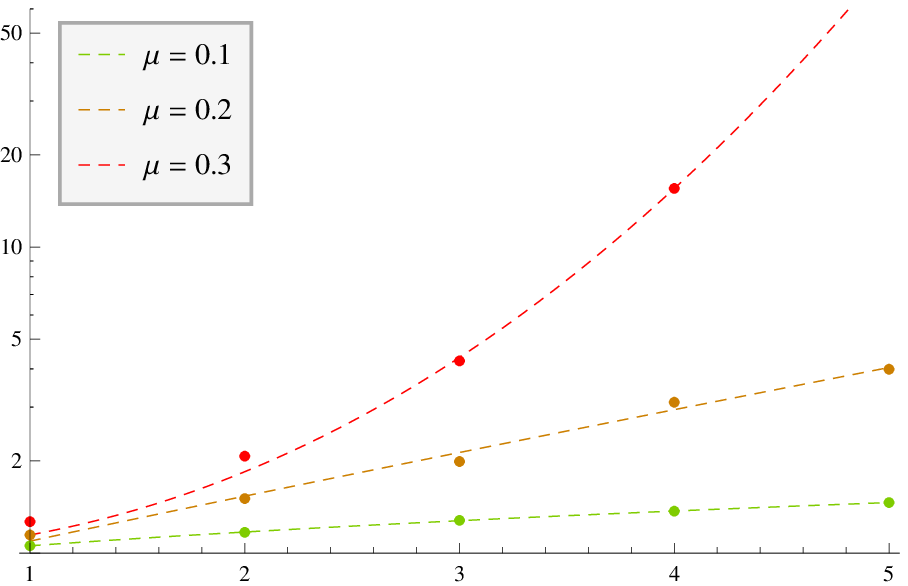} \quad \quad
\includegraphics[width=7cm]{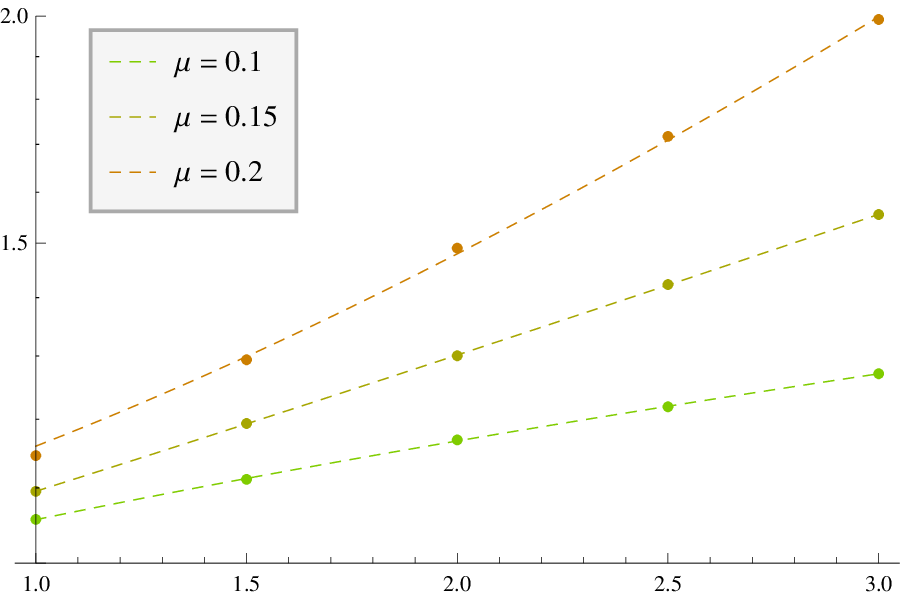}\\[5mm]
\caption{Solutions of Eq.~\eqref{eq: Hq} as function $q=q(\mu)$ with  constant time $t$ -- Fig.~a), resp. Fig.~b) (for small values of $\mu$)  -- or as function $q=q(\mu)$ with constant $t$ -- Fig.~c), resp. Fig.~d) (for small values of $t$). These solutions show, which $q$'s have the ability to detect non-trivial multifractal scaling in the multi-scale model defined in Sect.~\ref{Sec3b}. Obtained curves in a log-lin plots represent the best polynomial fits adapted to given points. The curves obtained imply that $q$ grows quickly --- faster than an exponential function of $\mu$, respective $t$. On the other hand, the exponential dependence
is well satisfies at small values of $t$.}
\label{fig: muq}
\begin{picture}(20,7)
\put(215,72){$t$}
\put(-10,72){$t$}
\put(215,224){$\mu$}
\put(-10,224){$\mu$}
\put(-210,205){$q$}
\put(10,205){$q$}
\put(-210,355){$q$}
\put(10,355){$q$}
\put(-221,295){a)}
\put(8,295){b)}
\put(8,145){d)}
\put(-221,145){c)}
\end{picture}
\end{center}
\end{figure}

Because $\int_{\mathds{R}} \ud x \   L_{1, \ \!\mu}(x)\ \! \ln L_{1, \ \!\mu}(x)$ is a {\em monotonically increasing} function of $\mu$ (see, e.g., Ref.~\cite{Johnson:04} --- the more leptocurtic $L_{1, \ \!\mu}$ the higher Shannon's entropy/ignorance), the derivative term in
(\ref{3.24aaa}) is positive. Since also $\ln t \geq 0$ (note that $t$ is an integer fulfilling $t \geq 1$), the condition (\ref{3.24aaa})
implies that $\mu'(s) = 0$ or equivalently that $\mu(s)$ is a constant. This confirms the former observation that with Shannon's differential entropy one can well quantify only {\em monofractal} time series. On the other hand, when $q \neq 1$ we can address also the case with a non-constant scaling exponent. This is possible because the expression within $[\ldots]$ {\em can be zero} for  $q>1$.
In fact, it is not difficult to check numerically that $\int_{\mathds{R}} \ud x \ L^q_{1, \ \!\mu}(x) $ is for $q < 1$ monotonically increasing function of $\mu$ while for $q>1$ it is not. This in turn allows (at least for some time horizons $t$) for a non-trivial dependence of $\mu$  on $q$. This is depicted on Fig.~\ref{fig: muq}.
%
Of course, the fact that $[\ldots] =0$ does not, per se,  preclude a possibility that $\mu$ is still a constant (as seen from (\ref{3.24aaa})).
If this would be the case, the time series would be a monofractal, which --- being a $q$-independent statement --- could be
verified by looking at few $H_q$'s with $q>1$. If monofractality would be indeed confirmed then further analyzed could done solely via Shannon's entropy. Whether monofractal or multifractal, from the log-lin plot in the ($t, H_q$) plane
one can extract the desired scaling $1/\mu(s) = \delta(s) = \delta(s(q)) \equiv \delta(q)$ (cf. Ref.~\cite{Renyi}).
By realizing that the considered (dimensionless) time variable $t$ is measured in units $\mathfrak{s}$, we can the logarithmic term in (\ref{3.21.abc}) write as
\begin{eqnarray}
-\frac{1}{\mu(s)}\ln \mathfrak{s}  \ + \   \frac{1}{\mu(s)}\ln t\, ,
\end{eqnarray}
where both $\mathfrak{s}$ and $t$ are now measured in (dimensionless) seconds. By comparing this with (\ref{RE1}) we
see that for sufficiently small time lags $\mathfrak{s}$ (much smaller than a typical auto-correlation time) the generalized
dimension $D(q) = \delta(q) = 1/\mu(q)$. Our toy model system is thus clearly indicative of the RE's important r\^{o}le in
multifractal spectrum analysis inasmuch as a particular choice of $q$ can ``highlight'' specific time scales $s$.

For general heterogeneous empirical time sequences, the question remains, if we are able to
evaluate the RE for distributions and
parameters of interest. For instance, for negative $q$'s the situation is
notoriously problematic, because for PDF's with unbounded support the integral
of $p^q(x)$ does not converge. Of course, we can assume a truncated (or
regulated) model, where we would specify {\em minimal} and {\em maximal} value
of the support. This helps formally with the integral convergence, nevertheless,
regarding the time dependence of the PDF, it will be very problematic to define
such time-dependent bounds properly, and what more, this most certainly will
affect the scaling behavior. In fact, from the information theoretical ground,
one should refrain from using  R\'{e}nyi's entropy with negative $q$'s. This is
because  reliability of a signal decoding (or information extraction) is
corrupted for negative $q$'s (see, e.g., Ref.~\cite{Renyi}). In the following we
will confine our attention mostly on positive valued $q$'s.

It should be finally stressed that in
contrast to the discrete case (\ref{RE1}), the differential RE
$H_q(t)$ is not(!) generally positive.
In particular, a distribution which is more confined than a unit volume has less
R\'{e}nyi's
entropy than the corresponding entropy of a uniform distribution over a unit
volume, and hence it yields a negative $H_q(t)$ (see, e.g., Ref.~\cite{Renyi}).

\section{Probability histogram estimation and its error\label{sec.4}}
As we have seen in the previous section, when trying to estimate the scaling exponent $\delta(q)$, one needs to specify the underlying probability distribution $p(x,t)$
in the form of a histogram for each fixed $t$. Multifractality is, however, a continuous property.
Any tool designed to validate the multifractal property of a time series faces difficulties implied by finite size and
discretization and any technique that is supposed to verify the multifractality involves some interpolation scheme which
can make it prone to some bias.
Particularly, the formation of the histogram from the (unknown) underlying $p(x)$ should be done with care
because in the RE-based analysis one needs not only all $\hat{p}_i$'s, but also their powers, i.e., $\hat{p}_i^q$'s for different values of $q$. In this section we evaluate the error when $p(x)$ is replaced by its histogram and outline what choice of the bin-size $h$ is optimal in order to minimize the error in the RE estimation. We begin with basic definition of histogram and its properties, then analyze the situation for different values of $q$ and finally derive explicit formulas for optimal bin-widths for different values of $q$.

\subsection{Basic properties of histograms\label{sec.4.1}}

Let assume that from the underlying PDF  $p(x)$ we generate (e.g., via sampling) a histogram $\hat{p}(x)$
with some inherent bin-width $h$, i.e.
\begin{equation}
p(x) \ \mapsto \ \hat{p}(x) \ = \ \sum_{i=1}^{n_B} \frac{\hat{p}_i}{h} \ \!\chi_i(x)\, = \frac{1}{N h} \sum_{i=-\infty}^{\infty} \nu_i \chi_i(x)\, ,
\end{equation}
where $\chi_i(x)$ is a characteristic function of the interval $\mathcal{I}_i = [x_{\min{}} + (i-1)h, x_{\min{}} + ih]$, $n_B$ is the number of bins, $\hat{p}_i$ are the actual sampling values at time $t$ and $\nu_i$ are sampled counts, so that $\hat{p}_i = \nu_i/N$. In the latter equality, we formally extended the sum over all $i \in \mathbb{Z}$, which enlarges the domain of the histogram to the whole real axis, i.e. below $x_{\min{}}$ and above $x_{\max{}}$.
Indeed for $i \not\in \{1,\dots,n_B\}$ is $\nu_i = 0$. We note further that there is a simple relationship between bin-width $h$ and number of bins $n_B$, namely
\begin{equation}
n_B \ = \ \left\lceil \frac{x_{\max{}} - x_{\min{}}}{h} \right\rceil \, ,
\end{equation}
where $\lceil x \rceil$ denotes the ceiling function (i.e., the smallest integer not less than $x$).
Depending on concrete realization of sampling, i.e., on the series $\{x_i\}_{i=1}^N$, frequencies $\nu_k$ acquire values from $1$ to $N$,
probability that the frequency is equal to $k$ is equal to binomial distribution after $N$ trial, each with probability $p_k$, where
$p_k = \int_{\mathcal{I}_k} p(x) \ud x$, and indeed, in the large $N$ limit $\hat{p}_k \rightarrow  p_k$. From this point of view the
histogram represents a class of possible approximations of underlying PDF, which we further denote as $\mathcal{H}$.
The class $\mathcal{H}$ of all histograms contains piecewise constant functions,
which in each interval of length $h$ acquire values from the set $\{1/N,2/N,\dots,1\}$, and to each such histogram is assigned a probability of occurrence (determined by probabilities of $\nu_k$). Construction of histogram from sampled data $\{x_1,\dots,x_N\}$
corresponds thus to a choice of a one particular histogram from the class $\mathcal{H}$. The $q$-th power of the histogram (necessary for the RE estimation) is then simply given as
\begin{equation}\label{eq: qhistogram}
\hat{p}^q(x) \ = \ \frac{1}{N^q h^q} \sum_{i=1}^{n_B} \nu_i^q \chi_i(x)\, ,
\end{equation}
which is a simple consequence of the definition of characteristic functions $\chi_i$.  Our aim, of course, is to create such histogram that would be the best representation of the underlying distribution, in a sense, that it would yield the smallest possible error
in the RE evaluation. For this purpose, it is necessary to find an appropriate measure between underlying PDF and sampled histogram that would be both computationally tractable and conceptually well motivated. The crucial question that should be then addressed is whether the optimal bin-width can be $q$-independent or whether the minimality of the error will require some non-trivial $q$-dependence.
Next subsections are devoted to the discussion of different measures used for measuring the distance (or dissimilarity) between underlying PDF and its histogram and to the ensuing $q$-dependence issue.

\subsection{Distance between histogram and probability distribution\label{sec.4.2}}

In statistical estimation problems, there exits a number of measures of dissimilarity between probability distributions, among these Hellinger coefficient, Jeffreys distance, Akaike's criterion, directed divergence, and its symmetrization $J$-divergence provide paradigmatic examples. In the context of the RE-based analysis, the most natural measure is the R\'{e}nyi information divergence of
$\hat{p}$ from $p$. This is defined as~\cite{Renyi:76}
\begin{eqnarray}
D_q(p||\hat{p}) \ = \ \frac{1}{q-1} \ln  \int_{\mathds{R}} \ud x \ \! \ \!\hat{p}^{1-q}(x) p^q(x) \, .
\label{4.2.28.aa}
\end{eqnarray}
Apart from a multiplicative prefactor, the divergence $D_q(p||\hat{p})$ coincides with
the Chernoff probabilistic distance of order $q$ (see, e.g., Ref.~\cite{Basseville:89}). From information theory it is known (see, e.g., Ref.~\cite{transferent,Renyi,Renyi:76}) that the  R\'{e}nyi information divergence represents a measure of the information lost when the
PDF $p(x)$ is replaced (or approximated) by the PDF  $\hat{p}(x)$.
Note that in the limit $q\rightarrow 1$ one recovers the usual Shannon entropy-based Kullback--Leibler divergence. In the case of histogram sampling, the {\em expected} R\'{e}nyi information divergence
\begin{equation}\label{eq: expectedRD}
\E_\mathcal{H}\left[D_q(p||\hat{p})\right] \ = \ \frac{1}{q-1} \E_\mathcal{H} \left[\ln  \int_{\mathds{R}} \ud x \ \! \ \!\hat{p}^{1-q}(x) p^q(x)\right] \, ,
\end{equation}
represents the {\em mean} loss of information. Symbol $\E_\mathcal{H}[\ldots]$ represents the ensemble average with respect
to the ensemble of histograms $\hat{p}(x)$.
Unfortunately, working with the measure (\ref{4.2.28.aa}) brings about some technical difficulties related to the non-linear structure of the RE.
This can be circumvented by progressing with other measures that are closely related to R\'{e}nyi informational divergence and yet are
computationally more tractable. Among these, we have found that the squared $L_2$-measure serves best our purposes.
The latter quantifies the distance between the underlying PDF and histogram as
\begin{equation}
\|p-\hat{p}\|_{L_2}^2 \ = \ \int_{\mathds{R}} \ud x \left[(p(x) - \hat{p}(x))^2\right]\, .
\end{equation}
This measure has number of desirable properties, as we shall see in the following subsections. The relation
between the R\'{e}nyi information divergence and $L_2$-measure is provided as follows: by using Jensen's inequality for the logarithm
\begin{eqnarray}
1 - \frac{1}{z} \ \leq \ \ln z \ \leq \ z-1 \, ,
\end{eqnarray}
(valid for any $z>0$), we obtain that
\begin{eqnarray}\label{eq: renyiL1}
|D_q(p||\hat{p})| \ \leq \ \frac{c_q}{|q-1|}\int_{\mathds{R}} \ud x \ |p^q(x) - \hat{p}^q(x)|\, ,
\label{4.24.ab}
\end{eqnarray}
where
\begin{eqnarray}
c_q \ = \ \max\left\{1, \left(\int_{\mathds{R}} \ud x \ \! \ \! \hat{p}^{1-q}(x) p^q(x)\right) ^{-1}   \right\}.
\end{eqnarray}
Eq.~(\ref{4.24.ab}) is the $q$-generalization of the Csisz\'{a}r--Kulback inequality~\cite{Csiszar:67} known from Shannon's information theory.
Note that for $q \geq 1$ we have $c_q = 1$.  This is because for $q \geq 1$ we can write
\begin{eqnarray}
\int_{\mathds{R}} \ud x \ \! \ \! \hat{p}^{1-q}(x) p^q(x) \ = \ \sum_k h \ \! \hat{p}^{1-q}(\xi_{_k}) p^q(\xi_{_k})
\ = \  \sum_k \hat{p}^{1-q}_{_k} p^q_{_k} \ = \ \sum_k  \hat{p}_{_k} \left(\frac{p_{_k}}{\hat{p}_{_k}}\right)^{q} \ \geq \
\left(\sum_k {p}_{_k}\right)^{q} \ = \ 1  \, ,
\label{4.27.a}
\end{eqnarray}
where the integrated probabilities are defined as
\begin{eqnarray}
p_{_k} \ = \ \int_{kh}^{(k+1)h} \ud x \ \! \ \! p(x) \ = \ h\ \! p(\xi_{_k})\, , \;\;\;\;\;\;\;
\hat{p}_{_k} \ = \ \int_{kh}^{(k+1)h} \ud x \ \! \ \! \hat{p}(x) \ = \ h \ \! \hat{p}(\xi_{_k})\, ,
\end{eqnarray}
and where $\xi_k$ denotes a point in the interval $[kh,(k+1)h]$.
The last inequality in (\ref{4.27.a}) results from Jensen's inequality for convex functions.

The situation with $q\in [0,1)$ is less trivial because
\begin{eqnarray}
\int_{\mathds{R}} \ud x \ \! \ \! \hat{p}^{1-q}(x) p^q(x) \ < \ 1\,  ,
\end{eqnarray}
due to concavity of $\left({p_{_k}}/{\hat{p}_{_k}}\right)^q$. A simple (but not
very stringent) majorization of $c_q$ can be found by using
\begin{eqnarray}
\int_{\mathds{R}} \ud x \ \! \ \! \hat{p}^{1-q}(x) p^q(x) \ = \  \sum_k \hat{p}^{1-q}_{_k} p^q_{_k} \ \geq \ \sum_k \hat{p}^{1-q}_{_k} p_{_k}
\ \geq \ \mbox{min}(\hat{p}^{1-q}_{_i}) \ = \ [\mbox{min}(\hat{p}_{_i})]^{1-q} \, ,
\end{eqnarray}
and hence $c_q \leq [\mbox{min}(\hat{p}_{_i})]^{q-1}$.
%
%
Finally, from previous expressions, it is easy to see that
\begin{eqnarray}
D_q(p||\hat{p})^2 \ \leq \ \frac{c_q^2}{(q-1)^2}\left(\int_{\mathds{R}} \ud x \ |p^q(x) - \hat{p}^q(x)|\right)^2
\ \leq \ \frac{c_q^2}{(q-1)^2} \int_{\mathds{R}} \ud x \ (\hat{p}^q(x) - {p}^q(x))^2 \, ,
\label{4.1.34.ab}
\end{eqnarray}
where the second inequality follows from H\"{o}lder's inequality~\cite{gradstein:a}. Thus, we can estimate the expected R\'{e}nyi information divergence via the  $L_2$-distance, because
\begin{eqnarray}\label{eq: renyil2}
\E_\mathcal{H}[D_q(p||\hat{p})^2] \ \leq \ \frac{c_q^2}{(q-1)^2} \ \E_\mathcal{H}\left[\int_{\mathds{R}} \ud x \ [(p^q(x) - \hat{p}^q(x))^2] \right] \ \equiv \ \mathcal{C}_q  \int_{\mathds{R}} \ud x \ \E_\mathcal{H}[(p^q(x) - \hat{p}^q(x))^2]\, .
\end{eqnarray}
A substantial advantage in using the expected squared $L_2$-distance, i.e.
\begin{equation}\label{eq: expectedl2}
\E_\mathcal{H} \|p^2-\hat{p}^2\|_{L_2}^2 = \int_\mathds{R} \ud x \ \E_\mathcal{H} \left[(p^q(x) -\hat{p}(x))^2\right]\, ,
\end{equation}
lies in the possibility of interchanging integral and expectation value $\E_\mathcal{H}$. In the case, when the expected value stands before
the integral, we have to calculate the expected value over all frequencies $\{v_k\}_{k=1}^{n_B}$, such that $v_k \in \{1,\dots,N\}$ and $\sum_{k=1}^{n_B} v_k = N$. After interchanging integration and $\E_\mathcal{H}$, the ensemble averaging acts on $\hat{p}$ only locally, i.e.,
the expected value is calculated
only over one particular frequency $v_k$, for which $\chi_k(x) =1$. This brings about a significant simplification in calculations. This is the main reason, why to prefer in practical calculations the $L_2$-norm.

Let us note that in the course of calculations we have also used the $L_1$-norm  (see, Eq. \eqref{eq: renyiL1}).
Indeed, some authors use the $L_1$-norm  to measure the distance between histogram and PDF (e.g., Ref.~\cite{Hall}).
In the following we will stick to the mean square distance mainly because of its
computational superiority --- $L_2$-norms are notoriously easier to work with than $L_1$-norms~\cite{Scott}.
Naturally, different measures can generally lead to different results, and from the previous discussion, it might seem that
perhaps other measures could be even more convenient to employ.
Nevertheless, discussions in Refs.~\cite{Hall,Scott,Scott2,Silverman} imply that in case of
histograms which can be regarded as a $1$-parameter class of step functions, one can
``reasonably'' well  assume that optimal bin-widths obtained from different measures will
not drastically differ among themselves. This is also the case of R\'{e}nyi divergence, $L_1$ and squared $L_2$-measures, among which the $L_2$-measure is definitely the most convenient.

\subsection{Dependence of bin-width on $q$\label{sec.4.3}}

A natural question arises, if it is necessary to estimate the bin-width for each $q$ separately, or whether
it is not simply enough to estimate the bin-width only for one ``referential'' $q$, e.g., for $q=1$, and then work with such a histogram for
all other $q$-cases. We will now briefly motivate the necessity of introduction for $q$-dependent bin-width. More thorough discussion will be presented in the section to follow. Let us have a series $\{x_1,\dots,x_N\}$ sampled from $p(x)$. Let us also have a histogram $\hat{p}$
estimated from the data, with a bin-width $h$, that makes $\hat{p}$ optimal among all histograms that can be obtained from sampled data
$\{x_i\}_{i=1}^N$ by changing the value of $h$. For further convenience we denote $\Delta(x) = p(x) - \hat{p}(x)$.
The squared $L_2$-distance between $q$-th powers of $p(x)$ and  $\hat{p}(x)$ appearing in Eq.~\eqref{eq: expectedl2}
can be further conveniently rewritten as
\begin{eqnarray}
\|p^q - \hat{p}^q\|^2_{L_2} \ &=& \int_\mathds{R} \ud x \left(p^q(x)- \hat{p}^q(x)\right)^2 \ = \  \ \int_{\mathds{R}} \ud x \ \left(p^q(x) - \frac{1}{h^q} \sum_{i=1}^{n_B} \hat{p}^q_i \chi_i(x)\right)^2 \nonumber \\[2mm]
&=& \ \int_{-\infty}^{x_{\min{}}} \ud x \ p^{2q}(x) \ + \ \sum_{i=1}^{n_B} \int_{\mathcal{I}_i} \ud x \
\left(p^q(x) - \left[\frac{\hat{p}_i}{h}\right]^q \right)^2 \ + \ \int_{x_{\max{}}}^{\infty} \ud x \ p^{2q}(x) \, .
\end{eqnarray}
Assuming that the error $\Delta(x)$ is for each $x$ sufficiently small, we may approximate $p^q(x)$ as
\begin{equation}
p(x)^q \ = \  \left[\frac{\hat{p}_i}{h}\right]^q \ +\ \left( \begin{array}{c} q \\ 1 \\ \end{array}\right) \left[\frac{\hat{p}_i}{h}\right]^{q-1} \Delta(x) \ + \ \mathcal{O}\left(\Delta^2(x)\right) \, ,
\end{equation}
and therefore
\begin{equation}
\sum_{i=1}^{n_B} \int_{\mathcal{I}_i} \ud x \left(p^q(x) -\left[\frac{\hat{p}_i}{h}\right]^q\right)^2 \ \
\stackrel{\mathcal{O}(\Delta^2)}{\approx}
\ \ q^2 \sum_{i=1}^{n_B} \left(\left[\frac{\hat{p}_i}{h}\right]^{2(q-1)} \Delta_i^2\right)\, ,
\end{equation}
where $\Delta_i^2 \equiv \int_{\mathcal{I}_i} \ud x \ (\Delta(x))^2$. Denoting
\begin{eqnarray}
\Delta_0^{2q}  \ \equiv \ \int_{-\infty}^{x_{\min{}}} \ud x \ p^{2q}(x)\;\;\;\; \mbox{and} \;\;\;\; \Delta_{n_B+1}^{2q} \ \equiv \ \int_{x_{\max{}}}^{\infty}
\ud x \ p^{2q}(x)\, ,
\end{eqnarray}
the total squared $L_2$-distance can be expressed as
\begin{equation}
\|p^q - \hat{p}^q \|^2_{L_2} \ \approx \ \Delta_0^{2q} \ + \ q^2 \sum_{i=1}^{n_B} \left(\left[\frac{\hat{p}_i}{h}\right]^{2(q-1)} \Delta^2_i\right) \ + \ \Delta_{n_B+1}^{2q} \ \equiv \ \Delta_0^{2q} \ + \mathfrak{S}^2_q + \ \Delta_{n_B+1}^{2q} \, .
\label{4.24.a}
\end{equation}

In the following we will confine our discussion only to the middle sum
$\mathfrak{S}^2_q$. This is because $\mathfrak{S}^2_q$ depends
only on the choice of the histogram and hence on the value of $h$, while expressions
$\Delta_0^{2q}$ and $\Delta_{n_B+1}^{2q}$ dependent more on the actual underlying
PDF. We shall note that with increasing $N$ the values
$x_{\min{}}$ and $x_{\max{}}$ get closer to the respective borders
of the support of $p(x)$. So for sufficiently large $N$ the outer errors can be
omitted, and the total $L_2$-error can be represented only by $\mathfrak{S}^2_q$.
The discussion of $\mathfrak{S}^2_q$ can be divided into three distinct situations, according to the value of $q$:

\begin{description}
  \item[{$\bf{q < 0:}$}] for negative values of $q$, the sum $\mathfrak{S}^2_q$
accentuates the error that is particularly pronounced for distributions with
extremely small probabilities $\hat{p}_i$. This can be partially compensated by
smaller bin-width. However, in case of extreme
distributions it is very hard to decide, whether the estimated probability is
only an inappropriate outlier, or a sign of presence of extreme events in the
system. This error is usually the more pronounced the more negative $q$ is.
Consequently, the estimation of RE for negative $q$'s is extremely sensitive (in
fact, RE is in this case unreliable information measure~\cite{Renyi}) and many
authors evaluate RE only for positive $q$'s.
  \item[{$\bf{0 < q < 1:}$}] for these values, the exponent $q-1$ is larger than $-1$. Even though the errors from small probabilities are again accentuated, the error is bounded, because $\hat{p}_i^q \leq 1$ for $q \in (0,1)$,  so the error is not as dramatic as in the first case.
  \item[{$\bf{q > 1:}$}] in this case is the error diminished, because the factor $\hat{p}_i^{2(q-1)}$ suppresses the error exponentially with $2q$. The pre-factor $q^2$ does not grow as fast, and therefore the error is reduced in this case. Against this suppression acts $h^{2(1-q)}$, which increases the error for small $h$. It is thus generally better in this case not to over-fit the histogram too much.
\end{description}

From the aforementioned it should be also evident that minimization of the local
squared error $(\hat{p}(x)-p(x))^2$, or the integrated squared error
$\int_{\mathcal{I}_k} \left(\hat{p}(x)-p(x)\right)^2 \ud x$ does not necessarily
minimize the error of $(\hat{p}^q(x)-p^q(x))^2$. This is because we have to create histograms with different ($q$-dependent) bin-widths, which generally does not minimize the $L_2$-distance between histogram $\hat{p}(x)$ and the underlying PDF $p(x)$.

\subsection{Optimal width of the histogram bin\label{sec: optwidth}}

As discussed above, the proper histogram formation from the underlying
distribution is a crucial step in the RE estimation.  The
issue at stake is thus to find the optimal bin-width $h^*_q$ that minimizes the R\'{e}nyi informational error
in Eq.~\eqref{eq: expectedRD} (or better the mean-squared $L_2$-distance \eqref{eq: expectedl2}) and hence it provides
the least biased empirical value for the RE.

There exist several approaches for optimal bin-width choices in the literature that can be well employed in our mathematical framework.
In this connection one
can mention, e.g., {\em Sturges rule}~\cite{Sturges}, that estimates the number
of bins of the histogram as $n_B = 1 + \log_2 N$, which is motivated by the
histogram of binomial distributions. This rule is very useful when visualizing
data, but in case of PDF approximations, one can find more
effective prescriptions. Along those lines, particularly suitable is the
classic {\em mean square error~}(MSE) method (see, e.g., Ref.~\cite{Lehmann})
which, among others, allows to quantify the difference between $\hat{p}^q(x)$ and a $q$-th power of the
underlying PDF in terms of previously discussed  mean-squared $L_2$-distance. This leads us to the task of solving
\begin{equation}\label{eq: minh}
\min_{h \in (0, \infty)} \int_\mathds{R} \ud x \, \E_\mathcal{H}\left[(p^q(x)-\hat{p}^q(x))^2\right] \ = \ \min_{h \in (0, \infty)} \sum_{k=1}^{n_B} \int_{\chi_k} \ud x \, \E_{\nu_k}\left[\left(p^q(x)- \frac{\nu_k^q}{h^q N^q}\right)^2\right]\, .
\end{equation}
Expression \eqref{eq: minh} implies that we minimize integrated expected local deviations between the underlying PDF $p^q(x)$ and  $q$-th power of it histogram (or simply \emph{$q$-histogram})
$\hat{p}^q(x)$.  We begin by first calculating $\E_\mathcal{H}[(p^q(x)-\hat{p}^q(x))^2]$ (for simplicity's sake we omit the subscript $\mathcal{H}$)
which can be conveniently rewritten as
\begin{equation}
\ \E[(\hat{p}^q(x) - p^q(x))^2] \ = \ \E[(\hat{p^q}(x)- \E[\hat{p^q}(x)])^2] \ + \ \E \left[(\E [\hat{p^q}(x)] - p^q(x))^2\right] \ = \ \mathrm{Var}(\hat{p}^q(x)) \ + \ \left[\mathrm{Bias}(\hat{p}^q(x))\right]^2\, .
\label{4.1.35.aa}
\end{equation}
The first term on the right-hand-side represents {\em local variance} of the estimator and the second term represents squared {\em local bias}
of the histogram, i.e.,
$\mathrm{Bias}(\hat{p}^q(x)) = \E[ \E[ \hat{p}^q(x) ] - p^q(x)]$. Eq.~(\ref{4.1.35.aa}) represents a local deviation of the $q$-histogram from the $q$-th power of underlying PDF, and so from the estimation theory point of view $\hat{p}^q(x)$ serves as an {\em estimator} of $p^q(x)$.

To progress with computation of $\mathrm{Var}(\hat{p}^q(x))$, we find from Eq.~\eqref{eq: qhistogram} that it corresponds to the calculation of variance of the quantity $\hat{p}^q(x) = {v_k^q}/{N^q h^q}$,
where $k$  labels the bin for which $\chi_k(x) = 1$. Naturally, $\nu_k$ is binomially distributed (as discussed in Sect.~\ref{sec.4.1}), namely $\nu_k \sim B(N,p_k)$, where $p_k$ is the integrated underlying probability adapted to $k$-th bin. Thus,
\begin{equation}
\mathrm{Var}(\hat{p}^q(x)) \ = \ \mathrm{Var}\left(\frac{\nu_k^q}{N^q h^q}\right) \ = \ \frac{1}{N^{2q} h^{2q}} \left(\E[\nu_k^{2q}]- \E[(\nu_k)^q]^2 \right) \, .
\end{equation}
This leads to calculation of fractional moments of binomial distribution, which is generally an intractable task, unless $q$ is natural. Indeed, when we have enough statistics, then the CLT implies (see, e.g., Refs.~\cite{Feller,Box}) that the binomial distribution
can be approximated by the normal distribution as $B(N,p) \sim \mathcal{N}(Np,Np(1-p))$. In this case we get
\begin{equation}
\E[\nu_k^q] \ \approx \ \int_\mathds{R} \ud z \ \! |z|^q \frac{1}{2 \pi N p_k (1-p_k)} \exp\left(- \frac{(z-(N p_k)^2)^2}{2N p_k (1-p_k)}\right) \, .
\end{equation}
Moment $\E[z^q]$ was replaced by the absolute moment $\E[|z|^q]$, because the latter value is real, while the first may not.
An integral can be done in closed form, namely
\begin{equation}\label{eq: qbinom}
\E[\nu_k^q] \ \approx \ \frac{1}{\sqrt{\pi}}\left(2 N p_k (1-p_k)\right)^{q/2}
\Gamma\left(\frac{1+q}{2}\right) \exp\left(-\frac{n p_k}{2(1-p_k)}\right)
{}_1 F_{1}\left(\frac{1+q}{2},\frac{1}{2} ; \frac{N p_k}{2(1-p_k)}\right)\, ,
\end{equation}
where ${}_1 F_{1}(\alpha,\beta;z)$ is a \emph{confluent hypergeometric function}~\cite{Ryzhik} defined as
\begin{equation}
{}_1 F_{1}(\alpha,\beta;z) \ = \ 1 + \frac{\alpha}{\beta \cdot 1!} z +\frac{\alpha(\alpha+1)}{\beta(\beta+1) 2!}z^2 + \dots \ = \ \sum_{j=0}^\infty \frac{(\alpha)_j}{(\beta_j) j!} z^j\, .
\end{equation}
Symbol $(\alpha)_k = \alpha (\alpha+1)\dots(\alpha+k)$ is the \emph{Pochhammer symbol}~\cite{Ryzhik}. According to Ref.~\cite{Lebedev},
the confluent hypergeometric function can be for sufficiently large $z$ asymptotically expanded as
\begin{equation}
{}_1 F_{1}(\alpha,\beta;z) \ = \ \frac{\Gamma(\beta)}{\Gamma(\alpha)} e^z z^{-(\beta-\alpha)}\left(1 + (\beta-\alpha)(1-\alpha)z^{-1} +\mathcal{O}(z^{-2})\right)\, .
\end{equation}
Reinserting this back into Eq.~\eqref{eq: qbinom} we obtain
\begin{equation}
\E[\nu_k^q] \ = \ N^q p_k^q \left(1 + \frac{1}{2}q(q-1) \frac{(1-p_k)}{N p_k} + \mathcal{O}(N^{-2})\right).
\end{equation}
Consequently, the leading order in $N$ yields the local variance
\begin{equation}
\mathrm{Var}\left(\hat{p}^q(x)\right) \ = \ \frac{1}{N^{2q}h^{2q}} N^{2q} p_k^{2q}\left[ q^2 \frac{1-p_k}{N p_k} + \mathcal{O}(N^{-2})\right] \ = \ \frac{q^2 p_k^{2q-1}(1-p_k)}{h^{2q} N} + \mathcal{O}(N^{-2}) \ \leq \ \frac{q^2 p_k^{2q-1}}{h^{2q} N}\ + \ \mathcal{O}(N^{-2}) \, .
\end{equation}
Similarly, the bias from Eq.~(\ref{4.1.35.aa}) can be cast in the form
\begin{equation}
\mathrm{Bias}(\hat{p}^q(x)) \ =  \ \frac{\E[\nu^q_k]}{N^q h^q}-p^q(x) \ = \ \left(\frac{p_k}{h}\right)^q -p^q(x) + \mathcal{O}(N^{-1}) \, .
\end{equation}

When we want to calculate the total error for all points of histogram, we should  integrate over all local errors and obtain the {\em mean integrated square error}~(MISE), which equals to
\begin{equation}
\mbox{MISE}(\hat{p}^q) \ \equiv \ \int_\mathds{R} \mathrm{Var} (\hat{p}^q(x)) \ud x \ + \ \int_\mathds{R} \left[\mathrm{Bias}(\hat{p}^q(x))\right]^2 \ud x \, .
\end{equation}
Because ultimately only the leading terms in $N$ are relevant, we can   consider the integrated variance in the form
\begin{equation}
\int_\mathds{R} \mathrm{Var} (\hat{p}^q(x)) \ud x \ = \ \sum_{k=-\infty}^{\infty} \int_{\mathcal{I}_k} \mathrm{Var} (\hat{p}^q(x)) \ud x  \ \approx \  \sum_{k=0}^{n_B+1} \frac{q^2 p_k^{2q-1}}{h^{2q-1} N}\, ,
\end{equation}
By applying further the mean value theorem for $p_k$, i.e.,
\begin{equation}
p_k \ = \ \int_{\mathcal{I}_k} p(x) \ud x \  = \ h p(\xi_k)\, ,
\end{equation}
we obtain that
\begin{equation}
\label{eq: ivariance}
\int_\mathds{R} \mathrm{Var} (\hat{p}^q(x)) \ \!\ud x \ \approx \ \sum_{k=-\infty}^{\infty} \frac{q^2 p_k^{2q-1}}{h^{2q-1} N} \ = \ \frac{q^2}{N h} \sum_{k=-\infty}^{\infty} p^{2q-1}(\xi_k) h \ \approx \ \frac{q^2}{N h} \int_\mathds{R} p^{2q-1}(x) \ \!\ud x\, .
\end{equation}
The leading $N$ behavior for the integrated squared bias can be obtained as follows.
We first write the integrated squared bias as a sum of integrated squared biases over $k$-th bin, i.e.,
\begin{equation}
\int_\mathds{R} \left[\mathrm{Bias} (\hat{p}^q(x))\right]^2 \ud x \ = \
\sum_{k=-\infty}^{\infty} \int_{\mathcal{I}_k} \left[\mathrm{Bias} (\hat{p}^q(x))\right]^2 \ud x\, .
\end{equation}
To proceed, let us look (without loss of generality) at a bin which lies between $0$ and $h$. We approximate the corresponding probability $p_{[0,h]}\equiv\int_0^h p(t) \ud t$ as
\begin{equation}
p_{[0,h]} \ = \ \int_0^h p(t) \ud t \ = \  \int_0^h \left(p(x) + (t-x) \frac{\ud p}{\ud x}(x)+ \dots\right) \ud t \ = \
h p(x) \ + \ h\left(\frac{h}{2} -x\right)\frac{\ud p(x)}{\ud x} \ + \ \mathcal{O}(h^3)\, .
\end{equation}
We note that because $x \in (0,h)$, the second term is of order $\mathcal{O}(h^2)$ and we can write the $q$-th power of $p_{[0,h]} $ as
\begin{equation}
p_{[0,h]}^q  \ = \ h^q p^q(x) \ + \  q h^{q-1} p^{q-1}(x) h\left(\frac{h}{2} -x\right)\frac{\ud p}{\ud x}(x) \ + \ \mathcal{O}(h^{q+2})\, .
\end{equation}
Consequently, the bias of that bin is to the leading order in $h$ equal to
\begin{eqnarray}
\int_0^h \left(\frac{h}{2} -x\right)^2 \left(q \frac{\ud p}{\ud x}(x) p^{q-1}(x)\right)^2 \ud x &=&  \int_0^h  \left[\left(\frac{h}{2} -x\right)\frac{\ud p^q}{\ud x}(x)\right]^2 \ud x \nonumber \\[2mm]
&=& \
\left(\frac{\ud p^q}{\ud x}(\xi_0)\right)^2 \int_0^h  \left(\frac{h}{2} -x\right)^2 \ud x \ = \ \frac{h^3}{12}\left(\frac{\ud p^q}{\ud x}(\xi_0)\right)^2.
\end{eqnarray}
The bias for other bins can be calculated in the same manner, and so finally we can write
\begin{equation}
\label{eq: ibias}
\int_\mathds{R} \left[\mathrm{Bias} (\hat{p}^q(x))\right]^2 \ud x \ \approx \ \frac{h^2}{12} \sum_{k=-\infty}^{\infty} \left(\frac{\ud p^q}{\ud x}(\xi_k)\right)^2 h
\ \approx \ \frac{h^2}{12} \int_\mathds{R} \left(\frac{\ud p^q}{\ud x}(x)\right)^2 \ud x\, .
\end{equation}
Combining Eq.~\eqref{eq: ivariance} with Eq.~\eqref{eq: ibias}, we get that the asymptotic (or leading-order; denoted as ``$\rm{l.o.}$'') mean integrated squared error~(AMISE) equals
\begin{eqnarray}
\label{4.1.29.a}
\mbox{AMISE}(\hat{p}^q) \ \stackrel{\rm{l.o.}}{\equiv} \  \E_\mathcal{H} \left[\int_\mathds{R} (\hat{p}^q(x)-p^q(x))^2\ud x\right]
= \  \frac{q^2}{N h} \int_\mathds{R} p^{2q-1}(x) \ \!\ud x \ + \  \frac{h^2}{12} \int_\mathds{R} \left(\frac{\ud p^q(x)}{\ud x}\right)^2\ud x\, .
\end{eqnarray}
In the limit $q \rightarrow 1$ we recover the classic result of Ref.~\cite{Scott}. Note that the bias increases when
$h$ increases while variance decreases when $h$ increases, i.e, we have to find a compromise between bias and variance
in order to be able to specify an optimal $h$. Minimization of the above AMISE function in $h$  gives the  optimal $h_q^*$ in the form
\begin{equation}
h_q^* \ = \ \sqrt[3]{\frac{6q^2}{N} \ \! \mathfrak{N}_q}\, ,
\end{equation}
where $\mathfrak{N}_q$ denotes
\begin{equation}\label{eq: Nq}
\mathfrak{N}_q \ = \ \frac{\int_\mathds{R} p^{2q-1}(x) \ \!\ud x}{\int_\mathds{R} \left({\ud p^q(x)}/{\ud x}\right)^2 \ud x}\, .
\end{equation}
By analogy with Scott~\cite{Scott} we can assume that the underlying PDF is the {\em normal distribution}
$\mathcal{N}(\mu,\sigma^2)$. In such a case one can write for $q > \frac{1}{2}$ that
\begin{equation}
\mathfrak{N}_q \ = \ \frac{4 \sqrt{\pi} \sigma^3}{\sqrt{q(2q-1)}}\, ,
\end{equation}
which gives the error
\begin{equation}
\mbox{AMISE}(\hat{p}^q) \ = \ \frac{q^2 \ \!(2 \pi)^{1-q}(\sigma^2)^{1-q}}{N h \sqrt{2q-1}} \ + \ \frac{h^2}{12} \sqrt{q}\ \!2^{-(1+q)} \pi^{-(1/2+q)}\sigma^{-(1+2q)}\, ,
\end{equation}
and the optimal bin-width $h^*_q$ factorizes as
\begin{equation}\label{eq: opt hq}
h^*_q \ = \ \sigma N^{-1/3} \sqrt[3]{24 \sqrt{\pi}} \frac{q^{1/2}}{\sqrt[6]{2q-1}} \ = \ h^*_1 \rho_q\, .
\end{equation}
Here $\rho_q = {q^{1/2}}/{\sqrt[6]{2q-1}}$ and $h^*_1$ is the optimal bin-width for $q=1$ (corresponding to classic result from histogram theory; see, e.g., Refs.~\cite{Scott,Silverman}). Note that from Eq.~\eqref{4.1.29.a}, the histogram  $\hat{p}^q$
converges in the mean square to $p^q(x)$ if $h^*_q \rightarrow 0$ and $Nh^*_q \approx N^{2/3} \rightarrow +\infty$.
This would be achieved when we could undertake more and more observations with smaller and smaller optimal bin-width,
but the optimal bin-width should not decrease too fast with $N$, namely the decrease should behave as
$h^*_q \approx N^{-1/3}$.

\begin{figure}[t]
\begin{center}
\includegraphics[width=6.5cm]{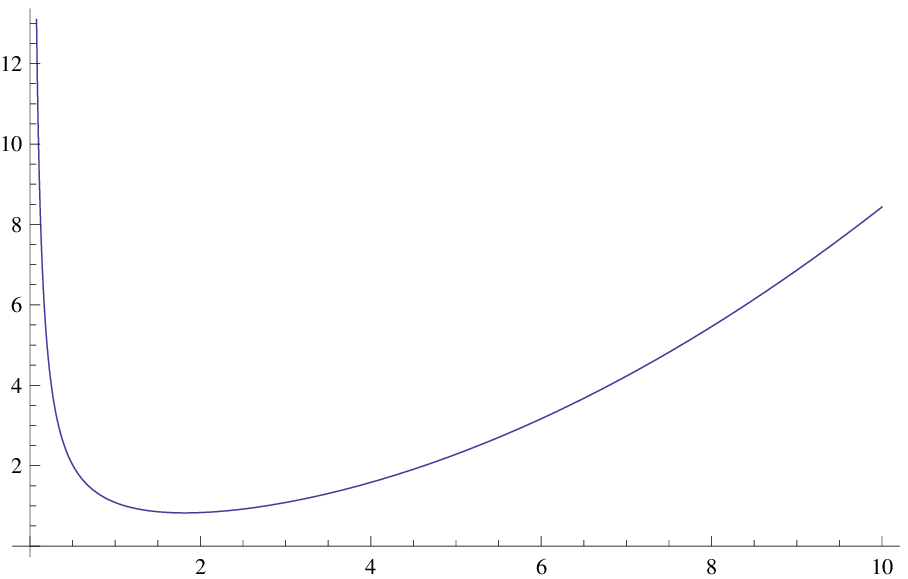} \qquad
\includegraphics[width=6.5cm]{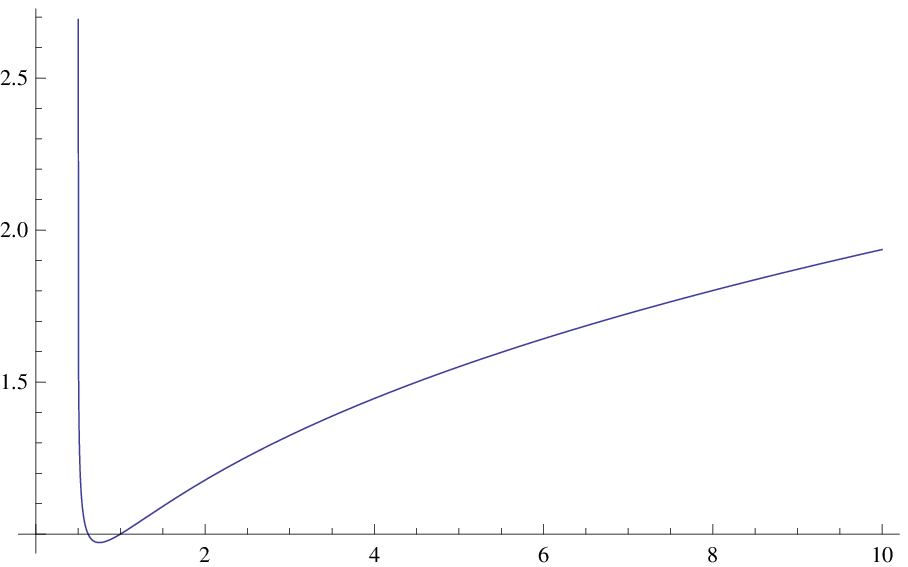}
\vspace{3mm}
\caption{From left: a) Shape of $\mbox{AMISE}(h,q=1) =\frac{1}{h} + \frac{h^2}{12}$ ($\sigma$ and $N$ are set so that the pre-fators are equal to one) as a function of $h$. We observe the optimal value $h^* = 6^{1/3} \doteq 1.817$ . b) Plot of  $\rho_q$ as a function of $q$.
For large $q$'s it goes as $q^{1/3}$, near the value $q=1/2$ it steeply diverges.}
\label{fig: amise}
\begin{picture}(20,7)
\put(-200,160){a)}
\put(5,160){b)}
\put(-220,100){\scriptsize{$\mbox{AMISE}(h)$}}
\put(10,100){$\rho_q$}
\put(-100,40){$h$}
\put(120,40){$q$}
\end{picture}
\end{center}
\end{figure}

In practical estimations, theoretical standard deviation is often replaced with
the empirical one, which gives the rule for bin-width $\hat{h}^{Sc}_q$ obtained via Scott's rule in the form
\begin{equation}
\hat{h}^{Sc}_q \ = \ 3.5 \hat{\sigma} N^{-1/3} \rho_q \, .
\end{equation}
Here $\hat{\sigma}$ is the estimated standard deviation of the
series. Moreover, Freedman and Diaconis (FD) in
Ref.~\cite{FreedmanDiaconis} proposed more robust rule where the
factor $3.5 \hat{\sigma}$ is replaced with  $2 \ \! \mbox{IQR}$.
Abbreviation $\mbox{IQR}$ stands for the {\em interquartile range};
$\mbox{IQR} = x_{_{0.75}} - x_{_{0.25}}$. Unfortunately, this
estimation is not precise, because
\begin{equation}
\mbox{IQR}\left(\mathcal{N}(\mu,\sigma^2)\right) \ = \ 2 \sqrt{2} \ \!\mathrm{erfc}^{(-1)}\left(1/2\right) \sigma \ \approx \ 1.349 \sigma\, ,
\end{equation}
(function $\mathrm{erfc}^{(-1)}(z)$ is the {\em inverse complementary error function}~\cite{Ryzhik}). Contrary to the original FD theory, we need to care in our case more about the over-fitting of histograms because the multiplication constant $\rho_q $ is now accentuated for small and large $q$'s.
In fact, we show in Fig.~\ref{fig: amise}~b) the shape of function $\rho_q$, which for large values of $q$ goes as $\rho_q \sim q^{1/3}$, but for values close to $q = {1}/{2}$, it grows very rapidly to infinity.
When we replace $\hat{\sigma}$ by $\mbox{IQR}$ more precisely, the resulting bin-width rule $\hat{h}^{FD}_q$ acquires the form
\begin{equation}
\hat{h}^{FD}_q \ = \ 2.6 (\widehat{\mbox{IQR}}) N^{-1/3} \rho_q \, .
\end{equation}
Case, when $q \leq \frac{1}{2}$ cannot be covered for the case of PDF's with unbounded support, because in that case
\begin{eqnarray}
\mathfrak{N}_{q=1/2} \ \sim \ \mu(\mathrm{supp}(p))\, ,
\end{eqnarray}
which is the measure of the support for $p(x)$. This situation could
be rectified, e.g., by assuming that the PDF has a finite support (i.e.,
it is a {\em bounded distribution}),
which evokes the discussion about proper estimation of bounds of the
distribution. One could be, in principle, tempted to
fix the situation by generalizing Scott's procedure in such a way that the normal
distribution would be exchanged for some other L\'{e}vy (stable)
distribution~\cite{Kolmogorov,Sato}). This hope is unfortunately false.
In fact, Scott's procedure based on heavy-tailed stable distributions
restricts the parameter $q$ even more above $1/2$. For instance, in the
case of the L\'{e}vy $\mu$-stable distributions which asymptotically decay as
\begin{equation}
p_\mu(x) \sim \frac{l_\mu}{|x|^{1+\mu}} \quad \mathrm{for} \quad |x|
\rightarrow +\infty\, ,
\end{equation}
($l_\mu$ is a $\mu$-dependent constant~\cite{Kolmogorov}) the
finiteness of $\int_\mathds{R} p^{2q-1}(x) \ud x$, implies that $\mathfrak{N}_q$ from
Eq.~\eqref{eq: Nq} has sense only for
\begin{equation}
q \ > \ \frac{1}{2} + \frac{1}{2(\mu+1)} \, .
\end{equation}
On the other hand, our discussion from Section~\ref{Sec3b} indicates that for the identification
of a multi-time scaling are important only RE's with $q\geq 1$.

\subsection{Optimal bin width of multiple histograms\label{sec.4.5}}

In order to address the multi-scale dynamics, we have to estimate  in the presented method simultaneously more probability distributions $p^q_s$ for more time lags $\{s_1, s_2,\dots,s_m\}$, all with the same bin
width\footnote{According to the introductory definition in Section~\ref{Sec2}, this corresponds to the situation when the involved dynamics is multi-scaled, and hence a typical time lag varies.  This is practically utilized by considering multiples of a basic time lag $\mathfrak{s}$. E.g., for the investigated series S\&P 500 the basic time lag $\mathfrak{s}$ considered is $1$~day.}. The choice of number and values of $\{s_i\}_{i=1}^m$ depends on the concrete problem at hand. It is necessary to take such values, for which we have enough statistics, otherwise when $s_i \sim N$, we can have problems with quantity estimations. In the most cases it is thus optimal to choose $s_i$ at least one order of magnitude smaller than $N$. In the algorithm listed in~\ref{sect: appendix}, the sequence $\{s_i\}_{i=1}^m$ is chosen as $\mathcal{S} = \{2^i\}_{i=2}^{\lfloor \log_2 N \rfloor -3}$. The choice of scales as a geometrical series is expedient in two ways. First, when the regression $H_q(s) \sim \ln s$ is performed, then points $\{\ln s_i\}_{i=1}^m$ are on log-lin plot distributed uniformly. Second, the asymptotic computational complexity decreases from original $\mathcal{O}(N^2)$ to $\mathcal{O}(N \log_2 N)$.

The optimization on multiple time scales can be done by minimizing the sum of all errors from all histograms and finding such $h^*_q$ that minimizes the {\em total} asymptotic mean square error (TAMISE), which is defined as
\begin{eqnarray}
\mbox{TAMISE}(\{\hat{p}^q_{s_1},\dots,\hat{p}^q_{s_m}\}) \ \equiv \ \sum_{i=1}^m \mbox{AMISE}(\hat{p}^q_{s_i}) \ = \
\sum_{i=1}^m \left(\frac{q^2 (2 \pi)^{1-q}\hat{\sigma}_{s_i}^{2(1-q)}}{N_{s_i} h \sqrt{2q-1}} + \frac{h^2}{12} \sqrt{q}2^{-(1+q)} \pi^{-(1/2+q)}\hat{\sigma}_{s_i}^{-(1+2q)} \right) .
\end{eqnarray}
With this, one can immediately cast $h^*_q$ in the form
\begin{equation}
h^*_{q} \ = \ (24 \sqrt{\pi})^{1/3} \rho_q \ \! \sqrt[3]{\frac{\sum_{i=1}^m {\sigma_{s_i}^{2(1-q)}}/{N_{s_i}}}{\sum_{i=1}^m \sigma_{s_i}^{-(1+2q)}} } \ \equiv \
(24 \sqrt{\pi})^{1/3} \rho_q \ \!\mathcal{N}^\sigma_{q,m} \, .
\end{equation}
The function $\mathcal{N}^\sigma_{q,m}$ represents the way, in which particular histograms contribute to the optimal bin-width. Again, following Scott~\cite{Scott} we can replace theoretical standard deviations by empirical $\hat{\sigma}_{s_i}$ and obtain
\begin{equation}\label{eq: hscott}
\hat{h}^{Sc}_q \ = \ 3.5 \ \! \rho_q \ \! \mathcal{N}^{\hat{\sigma}}_{q,m} \, .
\end{equation}
On the other hand, following the FD strategy~\cite{FreedmanDiaconis}, we replace estimated standard deviations by interquartile ranges, and consequently obtain
\begin{equation}\label{eq: hfd}
\hat{h}^{FD}_q \ = \ 2.6 \ \! \rho_q \ \! \mathcal{N}^{\widehat{IQR}}_{q,m} \, .
\end{equation}
The multiplicative constant was determined from the case, when $m=1$. Unfortunately, in case of multiple histograms, the formula does not have such a nice property as Eq.~\eqref{eq: opt hq}, i.e., it does not factorize into a product of $\rho_q$ and \mbox{$q$-independent} part.
In passing we may note that for $m=1$ we recover original form of Eq.~\eqref{eq: opt hq}.

In passing, we note that according to the shape of the AMISE error function displayed on Fig.~\ref{fig: amise}~a) for $q=1$, it is better to overestimate the number of bins in histogram (having a little bit more bins than optimal) rather than underestimate it. The error for underestimated histograms grows faster than in the case of overestimated ones. This is also the principal reason why the FD approach is working well for normal distributions, even though it estimates a bit higher number of bins than indicated by Scott's method.

%
\section{Numerical analysis of MF-DEA method and probability estimation \label{Sec.5}}

\begin{figure}[t]
\begin{center}
\includegraphics[width=13.5cm]{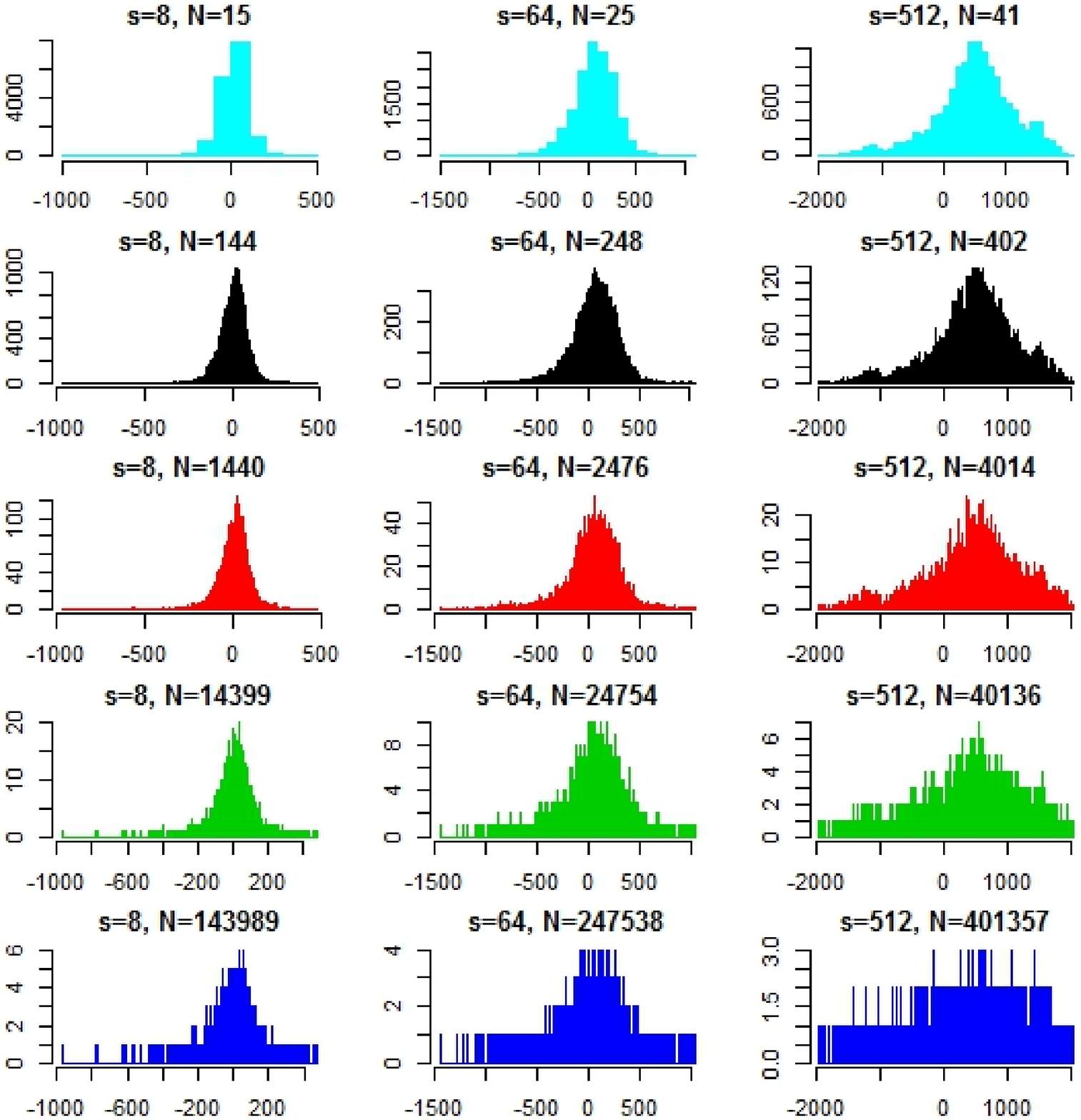}
\caption{Un-normalized (or frequency-based) histograms of the fluctuation sums $\sigma_s$, with $s = 8$, $64$ and $512$ and with bin-widths $h = 100, 10, 1, 0.1$ and $0.01$, measured in units $u=3\times 10^{-4}$ for better visualization. The optimal bin-width $h^*_{q=1}$  is specified in the table in Fig.~\ref{fig: opt}. We can see that far from the optimal value (listed in table under Fig.~\ref{fig: opt}), the shape of histogram is not appropriately approximating the theoretical probability distribution, i.e., we observe under-fitted or over-fitted histograms.}
\label{fig: hist}
\begin{picture}(20,7)
\put(-220,400){\footnotesize{$h=100$}}
\put(-220,325){\footnotesize{$h=10$}}
\put(-220,250){\footnotesize{$h=1$}}
\put(-220,175){\footnotesize{$h=0.1$}}
\put(-220,100){\footnotesize{$h=0.01$}}
\end{picture}
\end{center}
\end{figure}

\begin{figure}[t]
\begin{center}
\includegraphics[width=13.5cm]{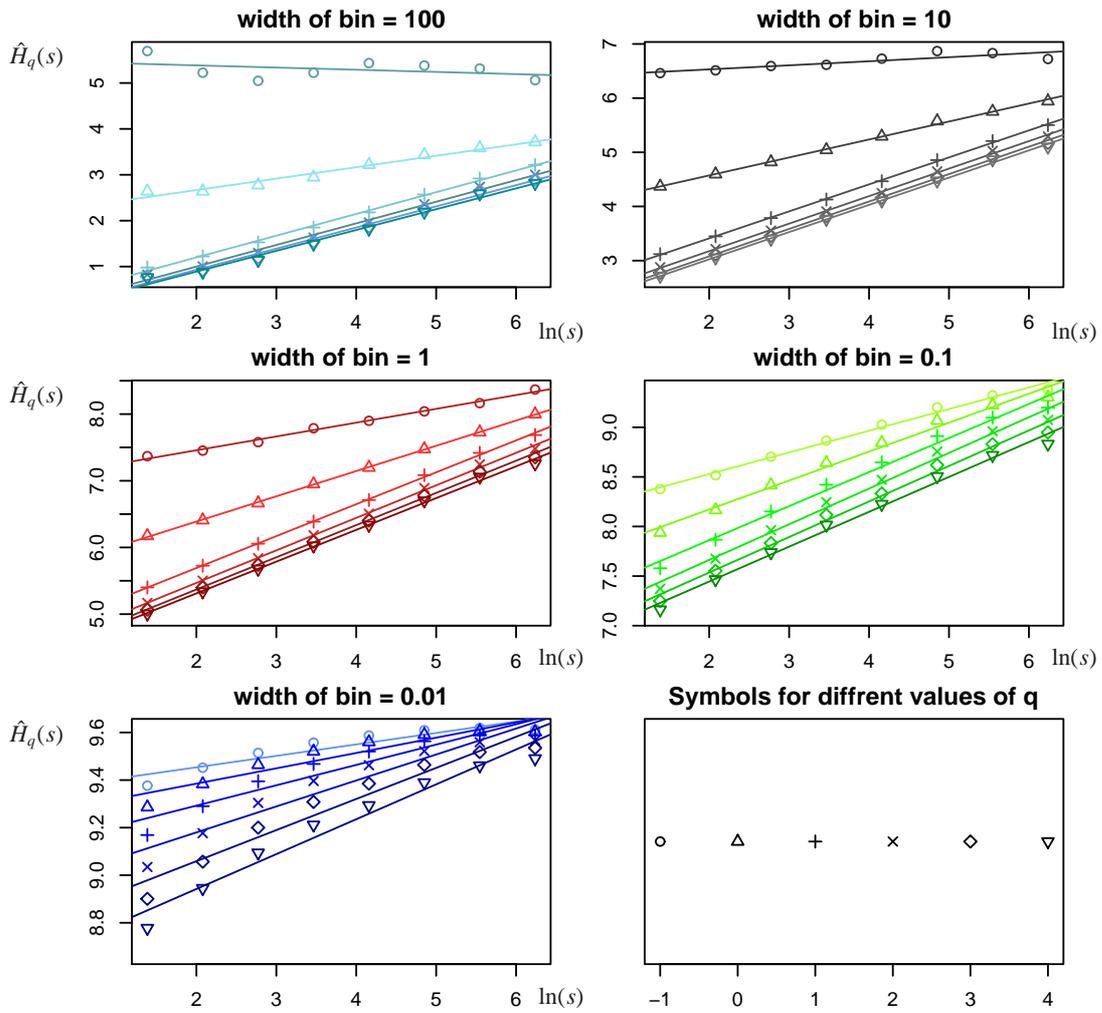}
\caption{Linear fits of estimated RE vs. $\ln s$ with $h=100,10,1,0.1,0.01$, measured in the same units as in Fig.~\ref{fig: hist}, i.e., in units $u=3 \times 10^{-4}$. Note, in particular, that the error is also non-trivially reflected in the defining linear regression for the scaling exponents (also dependent on $q$). This means that choice of a single $h^*$ valid for all $q$'s leads
to incorrect results and one has to resort to $h^*_q$ that is $q$-dependent.}
\label{fig: fit}
\begin{picture}(20,7)
\put(-210,408){\small{$\hat{H}_q(s)$}}
\put(-210,280){\small{$\hat{H}_q(s)$}}
\put(-210,152){\small{$\hat{H}_q(s)$}}
\put(-12,53){\small{$\ln(s)$}}
\put(-12,181){\small{$\ln(s)$}}
\put(-12,304){\small{$\ln(s)$}}
\put(180,181){\small{$\ln(s)$}}
\put(180,304){\small{$\ln(s)$}}
\end{picture}
\end{center}
\end{figure}

\begin{figure}[t]
\begin{center}
\includegraphics[width=14cm, height=6cm]{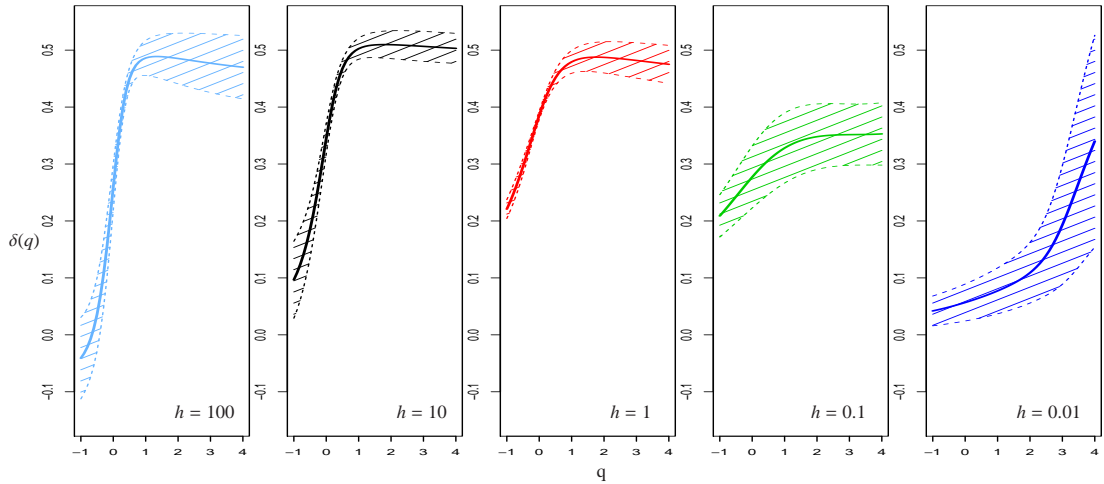}
\caption{Estimated $\delta$-spectra (middle line) and 99\% confidence intervals (shaded region) for different values of bin-width $h$. For bin-width far from the optimal width the spectrum is diminished
and confidence intervals get wider. Particularly, for under-fitted histograms the error is most dramatic for small $q$'s, on the other hand, for over-fitted histograms the error is most visible for large values of $q$'s.}
\label{fig: spec}
 \begin{picture}(20,7)
 \put(-138,65){\scriptsize{$h=100$}}
 \put(-55,65){\scriptsize{$h=10$}}
 \put(25,65){\scriptsize{$h=1$}}
 \put(100,65){\scriptsize{$h=0.1$}}
 \put(177,65){\scriptsize{$h=0.01$}}
 \put(-200,130){\scriptsize{$\delta(q)$}}
 \put(20,43){\scriptsize{q}}
 \end{picture}
\end{center}
\end{figure}

In order to illustrate the need for an accurate estimation of the empirical PDF  we  calculate the $\delta$-spectrum for different bin-widths and show that the error in the PDF estimation indeed substantially influences the multifractal spectrum. As an exemplary time series we select financial
time series of the stock index S\&P500 sampled at a daily rate during the period of time between January 1950 and March 2013 (roughly 16000 data points). Daily returns are visualized in Fig.~\ref{fig: fluct}~a). In the given time span the S\&P500 index can be considered as a good example of complex time series, because it exhibits a well patterned heterogeneous behavior. As demonstrated in Fig.~\ref{fig: hist},
the ensuing probability distribution was estimated for three different time lags and five distinct bin-widths. Note, in particular, that the histograms
which do not posses optimal bin-width are not approximating particularly well the underlying distribution. This is especially noticeable for widths far from the optimal
value $h^*_q$. In fact, the ensuing errors show up a non-trivial distribution when one is estimating regression
of scaling coefficients, cf. Fig.~\ref{fig: fit}. In cases, when the bin-width is not optimal, the estimated entropy does not exhibit a good linear behavior in log-lin scale. This is even accentuated for different choices of $q$. This error results in the fact that for different (non-optimal) choices of bin-widths, corresponding estimated $\delta(q)$-spectra are completely different (spectra are depicted in Fig.~\ref{fig: spec}).

Particularly, for extremely small bin-widths is the distribution disintegrated into simple (normalized) count functions of every element, because the probability that two or more values fall into the same bin-with $h \rightarrow 0$ tends to zero for a given constant length $N$. Let us note that the correct bin estimation is important also in the monofractal version of DEA, of Scafetta {\em at al.}~\cite{dea}.
Of course, if we estimate the spectrum within a small range of fluctuation times when $s_i \ll N$ for all $s_i$ (so $N \sim N_{s_i}$ and  $\sigma \sim s^H$, where $H$ is the Hurst exponent~\cite{Hurst}, usually $H \in (0,1)$), then we can estimate the optimal bin-width for the first histogram, and use it for other histograms as well. Nevertheless, for estimation across many scales or for evaluation of spectra at sensitive values of $q$, the choice of the proper $q$-dependent $h^{*}_q$ becomes increasingly important.
%

\subsection{Comparison of $\delta$-spectrum from different bin-width estimations}

\begin{figure}[t]
\begin{center}
\includegraphics[width=7.5cm]{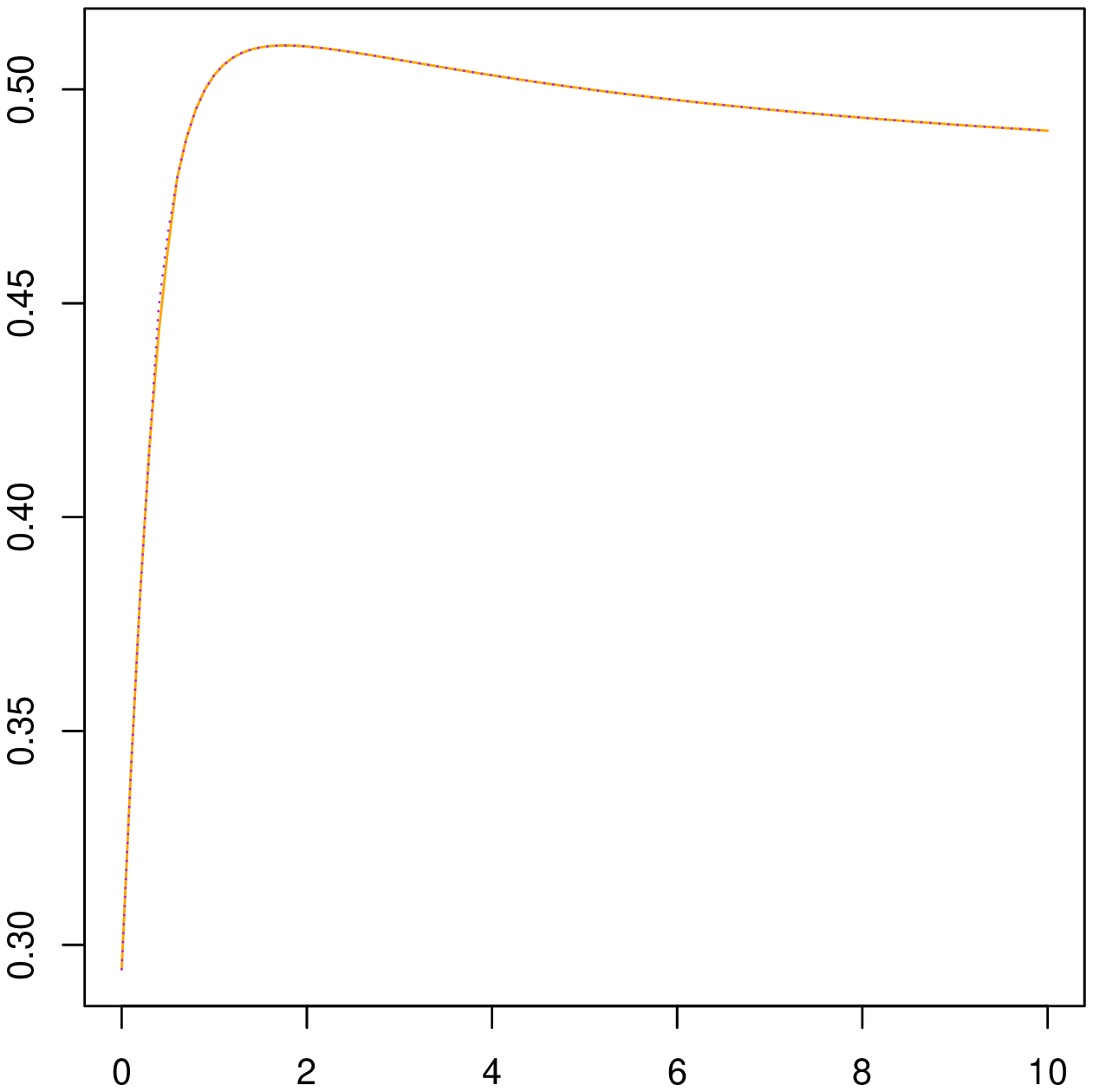} \
\includegraphics[width=7.5cm]{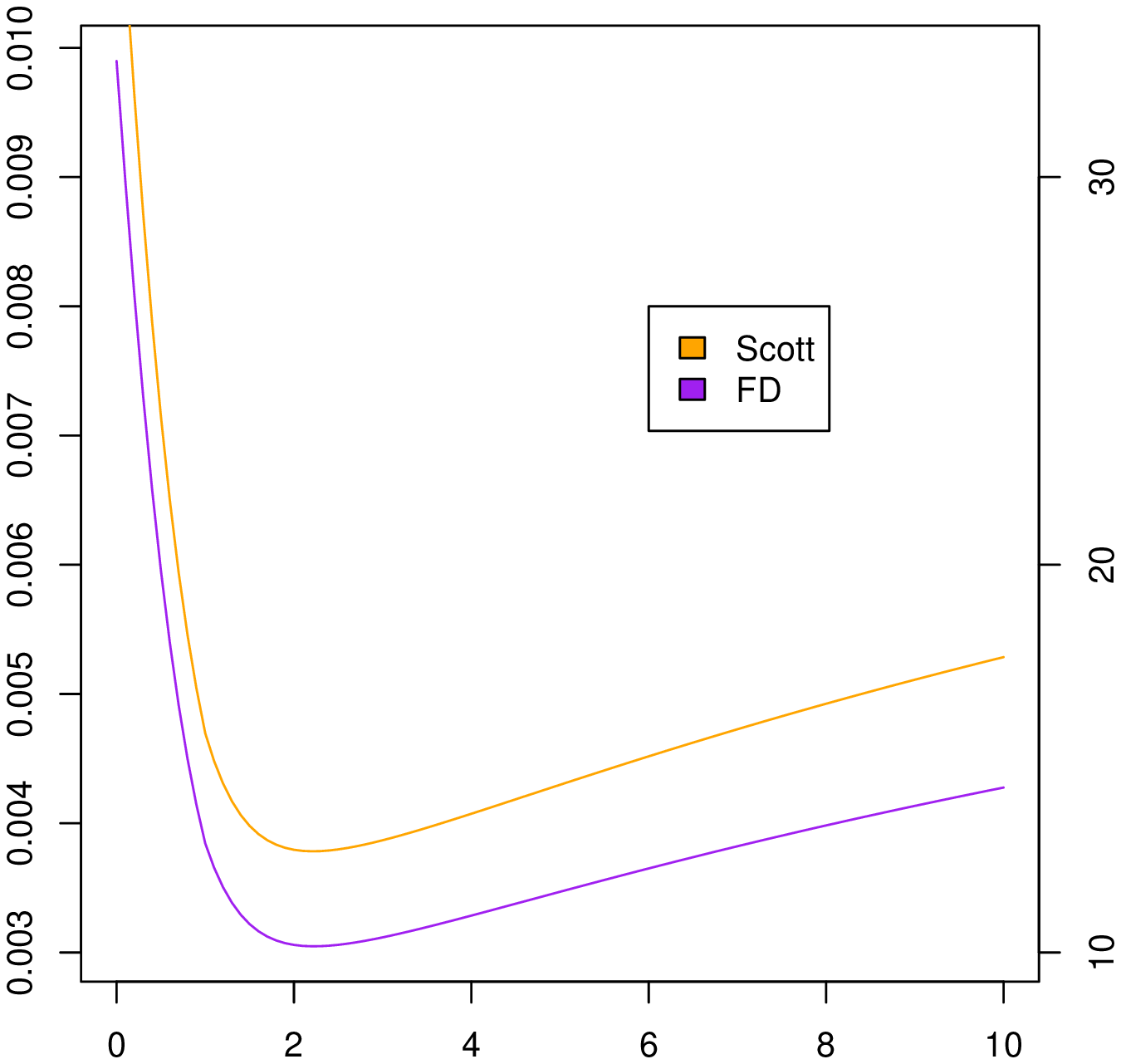}\\[5mm]
{\begin{tabular}{l|l|l}
\hline \hline
  method & optimal bin-width for $q=1$ & in multiples of $u=3\times 10^{-4}$\\
  \hline
  Scott &  \ 0.00470 & \ 14.10 \\
  Freedman--Diaconis & \ 0.00384 & \  12.81\\
  \hline \hline
\end{tabular}}
\caption{From left: a) $\delta(q)$ spectrum for bin-widths estimated by different methods. Spectra for both methods almost everywhere coincidence. b) Optimal bin-widths $\hat{h}^*_q$ for both methods. Left $y$-axis displays natural units, right $y$-axis compares the
width to multiples of $u=3 \times 10^{-4}$, for comparison with previous figures.
Under figures: Table with optimal values of $h^*_1$ for different methods, $h^*_q$ can be easily obtained from Eq.~\eqref{eq: opt hq}. The results are also converted to the same units like in Fig.~\ref{fig: hist}, so that the reader can easily compare the results with previous values.}
\label{fig: opt}
\begin{picture}(20,7)
\put(-110,114){$q$}
\put(-225,235){$\delta(q)$}
\put(110,114){$q$}
\put(-5,235){$h^*_q$}
\put(-220,310){a)}
\put(0,310){b)}
\end{picture}
\end{center}
\end{figure}

For every method discussed in Section~\ref{sec: optwidth}, we estimated the optimal bin-width and the spectrum. Results obtained are presented in Fig.~\ref{fig: opt} together with the table where the calculated optimal bin-widths are listed.
The figure implies the different spectrum for the two aforementioned approaches. We can observe
that even though the optimal bin-widths in Scott and FD method are different, the corresponding spectra can coalesce together in some cases. This can be caused by the fact, that financial
data are traded not for arbitrary price, which could be any real number, but prices that are always in dollars and cents (at the U.S. stocks) and the number expressed in dollars has maximally two digits after decimal point. This causes that the data are grouped at these price surfaces, and therefore different histograms can look completely the same, if these price surfaces fall into the same regions. In case of infinite precision of the empirical data (i.e., if the asset trading would be possible for any real-valued price), the spectra would be generally different.
Which of these methods is likely to be more efficient depends mainly on the concrete data-set, but, as we have argued in the
previous paragraph, it is generally better to overestimate the number of bins, than to underestimate it. Hence, from this standpoint is
the $q$-generalization of the FD approach more robust.

\section{Conclusions}

This paper has investigated the issue of the optimal bin-width choice for empirical probability histograms that typically appear in the framework of Monofractal and Multifractal Diffusion Entropy Analysis. With a simple model example of a heterogeneous time data sequence, we have demonstrated an important advantage in using the concept of differential R\'{e}nyi's entropy for study of time series steaming from multi-time-scale processes.  This is because among the most commonly used non-linear complexity measures, the RE-based fractal dimensions $\delta(q)$ and ensuing generalized dimensions $D(q)$ are directly associated to the way how PDF scales at respective time scales. In our model framework with
(non-Gaussian) stable-distributions with different stability coefficients for distinct time scales, we have seen that only RE's with $q \geq 1$ were of practical relevance. In addition, the model was tractable enough that we could extract the numerical dependencies of R\'{e}nyi's $q$ parameter on the local scaling $\mu$.

In order to be able to quantify in general time series scaling properties at different time scales we need the RE in a natural time. To this end we have considered fluctuation collection algorithm and ensuing diffusion RE. Such RE takes into account the sequential order of events, hence the RE obtained can be viewed as a dynamic entropy, i.e., it captures characteristics of the dynamics of a given system at respective scales. At the same time the important limitation of the fluctuation collection algorithm is its applicability to ergodic and Markovian systems.  Setting an appropriate generalization that would allow a natural passage between presented approach and non-ergodic stochastic processes (accelerating, correlated, path-dependent, or aging random walks) would be particularly desirable. Work along these lines based on the Fractional Brownian Motion is presently in progress.

Our subsequent investigation revealed that in order to obtain a reliable differential RE and the ensuing scaling exponents $\delta(q)$, the bin-width must be chosen with utmost care. We have argued that errors caused by estimation of the bin-width can be well quantified by the R\'{e}nyi information divergence (which is a $q$-generalization of more familiar Kullback--Leibler divergence) and by closely related $L_2$-distance. Obtained formulas for optimal bin-width reveal that for large values of $q$, the optimal bin-width grows approximately as $q^{1/3} h^*_1$, where $h^*_1$ is the optimal bin-width for $q=1$, which can be obtained from classic bin-width rules based on the $L_2$-distance (e.g., Scott or FD optimal
bin-width rule). On the other hand, our approach is, among others, based on asymptotic error evaluation which encounters a natural mathematical limit at $q\rightarrow \frac{1}{2}^+$ where the optimal bin-width begins to diverge. A root of this problem can be retraced to use on the normal distribution in classic AMISE approach, and can be (in principle) circumvented by assuming a bounded distribution, e.g. truncated Gaussian or L\'{e}vy distributions. Additional significant finding that emerged from this study is that aforementioned bin-width rules can be generalized straightforwardly for multiple histograms with the same bin-width. In this case, the resultant bin-width is derived from minimizing the total asymptotic mean integrated squared error. The rule for optimal bin-width of multiple histograms appears to be even more important, because optimal bin-widths on different scales and for different $q$ can markedly differ, and the rule prescribes, how to average these optimal bin-widths, in order to retain all errors on some acceptable level.

Presented analysis clearly indicated that the main advantage of MF-DEA approach lies in the robustness of the method, especially for large values of $q$, i.e., when $q \gg 1$. Some other previously discussed methods, mainly those, which are based on estimation of moments $\E[X^q(t)]$, tend to fail for data that exhibit long-range correlations, heavy tails, or other kinds of black swan-like events. Yet, these data occur relatively often in various complex systems and even a qualitative description of their behavior is currently in the spotlight of many practitioners.

Obtained results are particularly pertinent in empirical studies where scaling, self-similarity and multifractal properties of time series are at stake. Paradigmatic examples of such complexly patterned data sequences are empirical financial time series as exemplified, e.g., by stock returns or by volatility (both in developed and emerging markets).  Our numerical analysis of the S\&P500 daily returns  gathered over the period from $1950$ till $2013$  revealed that the appropriate $q$-generalizations of Scott and FD rules represent a suitable and experimentally viable way for estimating multifractal scaling exponent $\delta(q)$ in a number of empirical financial time series. Apart from a faster attainment of the desired scaling, important advantage of the whole presented procedure resides in the fact that it can be summarized into a compact algorithm (written in the $R$ code, which is part of this paper) that can be subsequently used for efficient estimation of the $\delta(q)$-spectrum (and ensuing generalized dimension $D(q)$)  of complex long-term  data-sets.


\section{Acknowledgements}

A particular thank goes to X.~Sailer of Nomura, Ltd., for helping us with the financial data. We are grateful also for comments from H.~Kleinert and H.~Lavi\v{c}ka which have helped us to understand better the ideas discussed in this paper. This work was supported by the Grant Agency of the Czech Republic, Grant No. GCP402/12/J077 and the Grant Agency of the Czech Technical University in Prague, Grant No. SGS13/217/OHK4/3T/14.


\appendix
\section{Source code of MF-DEA algorithm in programming language R}\label{sect: appendix}
\begin{verbatim}
#install.packages('GLDEX')
require(GLDEX)

spectrum <- function(X, method = 'FD'){
############################
# Input parameters
############################

if(length(X) < 128) return('Not enough data')
scale <- 2^seq(2:floor(log(length(X),2)-3))      #set of time lags
q <- seq(0,10, by= 0.1)                          #set of q's

#############################
# Fluctuation collection
#############################

fluctuation <- matrix(ncol=length(X), nrow=length(scale))
for(t in 1:length(scale)){
  for(s in 1:(length(X)-scale[[t]])){
    fluctuation[t,s] <- sum(X[s:(s+scale[[t]]-1)])
  }
}

#############################
# Estimation of hstar
#############################

sigmaL <- array(dim = length(scale))
for(i in 1:length(scale)){
  sigmaL[[i]] <- sd(na.omit(fluctuation[i,]))
}
sigma <<- sigmaL

IQRangeL <- array(dim = length(scale))
for(i in 1:length(scale)){
  IQRangeL[[i]] <- IQR(na.omit(fluctuation[i,]))
}
IQRange<<- IQRangeL

lengthsL <- array(dim = length(scale))
for(i in 1:length(scale)){
  lengthsL[[i]] <- length(na.omit(fluctuation[i,]))
}
lengths<<- lengthsL

rho <- function(q){
if(q>1){
return(q^(1/2)/(2*q-1)^(1/6))}
else{
return(1)}
}

#########   Scott hstar  ###########
hstarS <- function(q){
return(rho(q)*3.5*(sum ( sigma^(2*(1-q)) /lengths ) / sum( 1/ (sigma^(1+2*q)) ))^(1/3))
}

###### Freedman-Diaconis hstar  ####
hstarF <- function(q){
return(rho(q)*2.6*(sum ( IQRange^(2*(1-q)) /lengths ) / sum( 1/ (IQRange^(1+2*q)) ))^(1/3))
}

#########  Choice of method  #######
if(method == 'Scott'){
  hstar <- hstarS}
else{
  hstar <- hstarF
}

#############################
# Estimation of histogram
#############################

Sq <- matrix(nrow = length(q),ncol = length(scale))
tauq <- matrix(nrow = length(q), ncol = 2)
for(n in 1:(length(scale))){
  for(i in 1:length(q)){
    p <- na.omit(fluctuation[n,])
    h <-hist(p, breaks = floor((max(p)-min(p))/hstar(q[[i]]))+1, plot = FALSE)
    pr <-fun.zero.omit(h$counts/length(p))

#############################
# Estimation of Renyi entropy
#############################

    if(q[i] == 1){
      Sq[i,n] <- -sum(pr*log(pr))}
    else{
      Sq[i,n] <- 1/(1-q[i])*log(sum(pr^q[i]))
    }
  }
}

for(i in 1:length(q)){
  fit <- as.vector(na.omit(Sq[i,]))
  model <- lm(fit ~ log(scale))
  tauq[i,] <- coefficients(model)
}
tq <- tauq[,2]
ret <- data.frame(q,tq)
return(ret)

} #end of function

\end{verbatim}





\bibliographystyle{elsarticle-num}
\bibliography{lit}






%
\end{document}